\documentclass[%
 reprint,
superscriptaddress,
showpacs,preprintnumbers,
amsmath,amssymb,
aps,
prb,
]{revtex4-2}
 
\usepackage{graphicx}
\usepackage{dcolumn}
\usepackage{bm}
\usepackage{hyperref}
\hypersetup{
    colorlinks=true,    
    linkcolor=blue,     
    citecolor=blue,     
    filecolor=blue,     
    urlcolor=blue       
}

\begin{document}


\title{Electro-optically tunable second-harmonic generation in lithium niobate metasurfaces boosted by Brillouin zone folding induced bound states in the continuum}

\author{Huifu Qiu}
\affiliation{Institute for Advanced Study, Nanchang University, Nanchang 330031, China}

\author{Xu Tu}
\affiliation{Institute for Advanced Study, Nanchang University, Nanchang 330031, China}

\author{Meibao Qin}
\email{qinmb@ncpu.edu.cn}
\affiliation{School of Education, Nanchang Institute of Science and Technology, Nanchang 330108, China}

\author{Feng Wu}
\affiliation{School of Optoelectronic Engineering, Guangdong Polytechnic Normal University, Guangzhou 510665, China}

\author{Tingting Liu}
\affiliation{School of Information Engineering, Nanchang University, Nanchang 330031, China}
\affiliation{Institute for Advanced Study, Nanchang University, Nanchang 330031, China}

\author{Shuyuan Xiao}
\email{syxiao@ncu.edu.cn}
\affiliation{School of Information Engineering, Nanchang University, Nanchang 330031, China}
\affiliation{Institute for Advanced Study, Nanchang University, Nanchang 330031, China}

\begin{abstract}
	
Lithium niobate (LN) metasurfaces exhibit remarkable Pockels effect-driven electro-optic tunability, enabling dynamic control of optical responses through external electric fields. When combined with their high second-order nonlinear susceptibility ($\chi^{2}$), this tunability is projected into the nonlinear landscape, realizing second-harmonic generation (SHG)-dominated functionalities in integrated photonics. However, achieving deep SHG modulation in LN metasurfaces remains challenging due to LN's limited refractive index tunability under practical driving voltage. To address this, we design an air-hole-structured LN metasurface by strategically adjusting air hole positions to induce Brillouin zone folding-enabled bound states in the continuum with ultrahigh quality factors ($Q>10^{4}$). Numerical simulations demonstrate a $2.59\%$ SHG conversion efficiency at 0.1 MW/cm$^{2}$ excitation and a modulation depth exceeding 0.99 under 15 V peak-to-peak voltage ($\Delta V_{\text{pp}}$). This work establishes a compact framework for electrically tunable nonlinear optics, advancing applications in integrated quantum light sources and programmable photonic chips.

\end{abstract}

\maketitle


\section{\label{sec:level1}Introduction}

The development of optical metasurfaces has marked a transformative advancement in materials science and optical technologies, providing unprecedented possibilities for extending the functionalities of conventional optical devices\cite{Kildishev2013, Yu2014}. As two-dimensional artificial nanostructures, metasurfaces enable precise control over light's phase, amplitude, polarization, and frequency through subwavelength-scale unit cell engineering\cite{So2023, Yao2023, Kuznetsov2024}. These capabilities have driven their widespread applications in cutting-edge fields such as metalenses\cite{Hu2023, Hou2025}, optical holography\cite{Li2020, Meng2025}, polarization manipulation\cite{Ding2020, Wang2021a}, and nonlinear optics\cite{Kruk2022}. However, the optical responses of conventional metasurfaces remain fixed post-fabrication due to their structural determinism, severely limiting their flexibility in dynamic optical field modulation and full-wavefront control. To overcome this limitation, tunable and reconfigurable metasurfaces have emerged as a critical research frontier\cite{Xiao2020}. Among these, electro-optically tunable metasurfaces—capable of actively modulating optical properties via external stimuli such as electric fields—stand out as a pivotal branch. Current technical approaches for electro-optic tuning include: mechanically reconfigurable systems\cite{Herle2023, Ding2024}, electrochemically driven systems\cite{EavesRathert2022, Kovalik2024}, phase transition systems using phase-change materials\cite{Lu2021, Quan2025} and liquid crystals\cite{Zou2019, Beddoe2025}, and field-effect systems\cite{Zhang2023a, BeneaChelmus2022, Zheng2024}. These innovations have significantly advanced metasurface reconfigurability, paving the way for next-generation adaptive optical devices, dynamic light-field manipulation platforms, and on-chip photonic systems.

Lithium niobate (LN), a high-performance electro-optic material, derives its optical tunability from the linear Pockels effect—a non-centrosymmetric crystal property where the refractive index linearly responds to an external electric field\cite{Chen2022, Fedotova2022, Boes2023}. This effect originates from the displacement of ions in LN's crystal lattice under an electric field, inducing a polarization-dependent refractive index modulation described by the electro-optic tensor. Moreover, LN's exceptionally high second-order nonlinear susceptibility ($\chi^{2}$) makes it an ideal candidate for second harmonic generation (SHG) in metasurface architectures\cite{Fedotova2020, Yuan2021, Ma2021a, Huang2022, Qu2022, Qu2023, Jiang2024, Qu2025}. Recent advances have validated the feasibility of dynamically controlling SHG in electro-optically tunable LN metasurfaces\cite{He2024, DiFrancescantonio2024, Kanyang2025}. However, the weak refractive index tunability of LN—governed by the linear Pockels effect—results in a limited modulation amplitude (typically $\sim10^{-3}$ at metasurface scales), thereby constraining the SHG modulation depth. To address this, achieving efficient SHG switching under practical driving voltages necessitates the integration of ultrahigh-quality-factor ($Q>10^{4}$) resonant modes with extremely narrow spectral linewidths.

Bound states in the continuum (BIC), as theoretically non-radiating dark modes with infinite $Q$-factors, exhibit unique advantages for light confinement\cite{Hsu2016, Hwang2022, Huang2023}. These modes enable exceptional electric field localization and are pivotal for enhancing nonlinear optical responses in metasurfaces\cite{Liu2019, Xu2019, Koshelev2019, Zheng2022, Liu2023a, Liu2023, Feng2023, Tu2024, He2024a, Liu2025, Sun2025}. The $Q$-factor of BIC depends inversely on the resonance linewidth ($Q\propto1/\Delta\lambda$)\cite{Koshelev2018}, where $\Delta\lambda$ is the full width at half maximum (FWHM). This inverse proportionality allows BIC to achieve ultrahigh $Q$-factors ($Q>10^{4}$) and extremely narrow spectral linewidths, forming the foundation for efficient nonlinear modulation. In particular, a special class of BIC—Brillouin zone folding-induced BIC (BZF-BIC)—emerges from periodic structural perturbations that fold Brillouin zones, coupling guided modes at non-$\Gamma$ points into the light cone for strong field localization\cite{Wang2023, Adi2024, Wu2024, Sun2024, Yue2025}. Recent studies demonstrate BZF-BIC's superior performance in nonlinear frequency conversion, achieving enhanced third-harmonic generation and four-wave mixing\cite{Qin2025}. Unlike conventional symmetry-protected BIC whose $Q$-factors rapidly decay in $k$ space, BZF-BIC maintains high $Q$-factors ($Q>10^{3}$) over broad wave vector ranges through zone folding, significantly boosting light-matter interactions. Furthermore, BZF-BIC exhibits enhanced structural robustness: its high $Q$-factor shows strong tolerance to fabrication-induced disorder, retaining orders of magnitude higher than symmetry-protected BIC under identical perturbations\cite{Wang2023}.

In this work, to achieve efficient modulation of SHG in LN metasurfaces, we synergistically integrate the superior electro-optic tunable nonlinear properties of LN with the unique advantages of BZF-BIC. The metasurface design comprises a unit cell of periodically patterned air holes in a LN thin film, featuring $C_{4v}$ group symmetry. By precisely adjusting the relative positions of air holes to introduce periodic structural perturbations, we modified the metasurface unit cell configuration, thereby inducing BZF-BIC with high $Q$-factor ($Q>10^{4}$) and ultra-narrow resonance linewidths. Numerical simulations demonstrate a $2.59\%$ SHG conversion efficiency at an input pump intensity of 0.1 MW/cm$^{2}$. Furthermore, on-to-off SHG switching (modulation depth $>0.99$) is achieved under a peak-to-peak modulation voltage of 15 V. This work provides a paradigm for designing nonlinear optical modulators, tunable optical filters, and all-optical switches based on metasurfaces, significantly advancing the development of high-efficiency integrated dynamic nonlinear photonic systems.

\section{Results and discussion}

\subsection{BZF-BIC in LN metasurfaces}

\begin{figure}[htbp]
	\centering
	\includegraphics[width=0.5\textwidth]{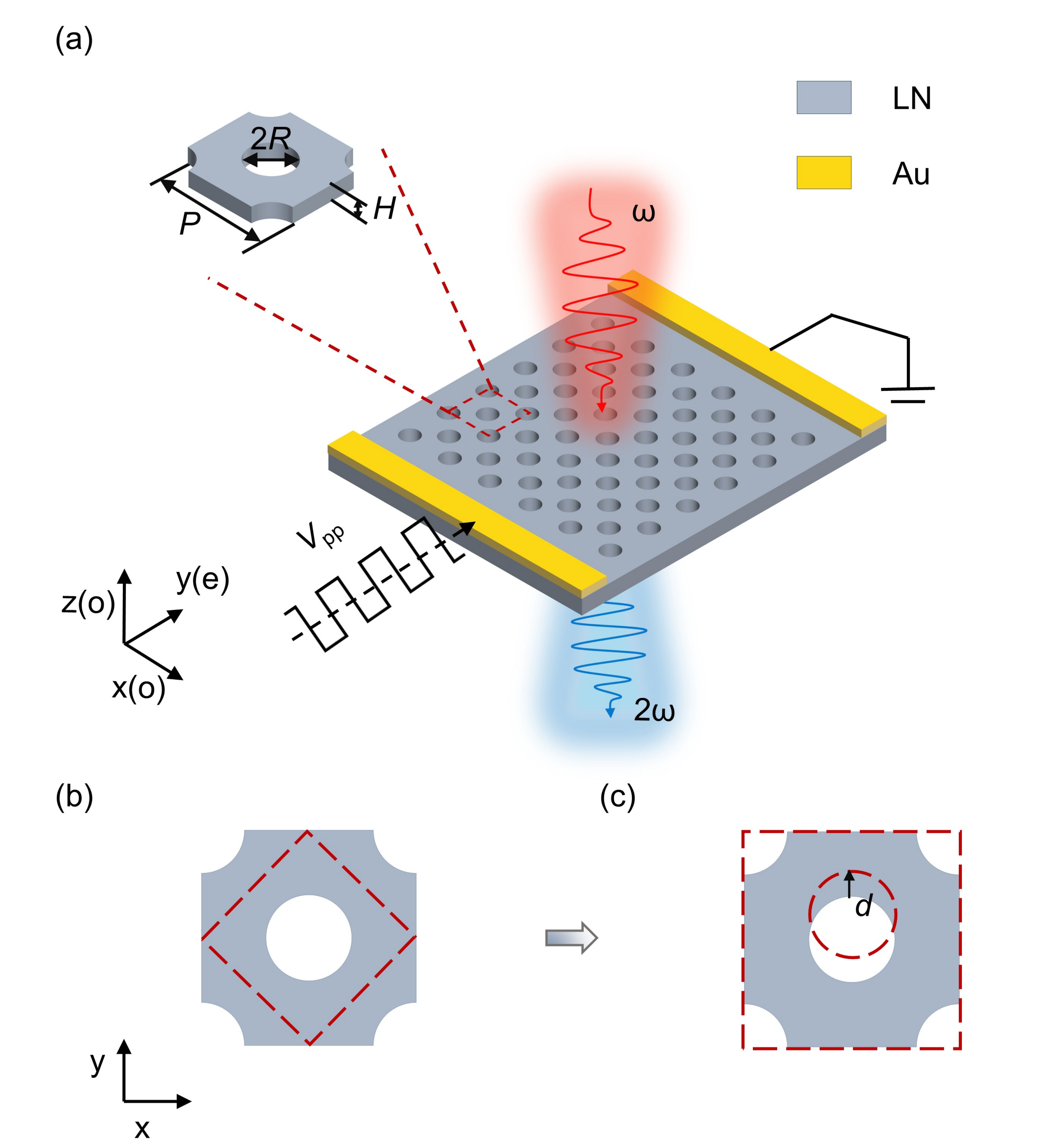}  
	\caption{(a) Schematic of the electro-optically tunable metasurface composed of an LN thin film patterned with a periodic array of circular air holes, with fixed geometric parameters: lattice constant $P=1000$ nm, thickness $H=300$ nm, and air hole radius $R=200$ nm. (b) Unit cell structure in the single-hole configuration, where the side length is $P/\sqrt{2}$. (c) Unit cell structure in the double-hole configuration, with $d$ defined as the displacement of the central air hole along the $y$-axis to introduce periodic perturbations.}
	\label{Fig1}
\end{figure}

As illustrated in Fig. \hyperref[Fig1]{1}(a), the proposed electro-optically tunable LN metasurface consists of a $x$-cut suspended LN thin film patterned with a periodic array of circular air holes, integrated with a pair of Au electrodes along the $y$-direction to enable electrical modulation. The metasurface adopts a unit cell with precisely defined geometric parameters: lattice constant $P=1000$ nm, thickness $H=300$ nm, and circular air holes of radius $R=200$ nm. Figs. \hyperref[Fig1]{1}(b) and \hyperref[Fig1]{1}(c) present the unit cell configurations before and after BZF. The original single-hole structure (pre-BZF) contains a single central circular air hole in each unit cell. To facilitate the excitation of BZF-BIC, we introduce periodic perturbations by displacing the central air hole along the $y$-axis. This structural modification transforms the unit cell into a double-hole configuration, consisting of a central circular air hole and four quarter-circular air holes etched at each corner of the square lattice.

\begin{figure*}[htbp]
	\centering
	\includegraphics[width=0.75\textwidth]{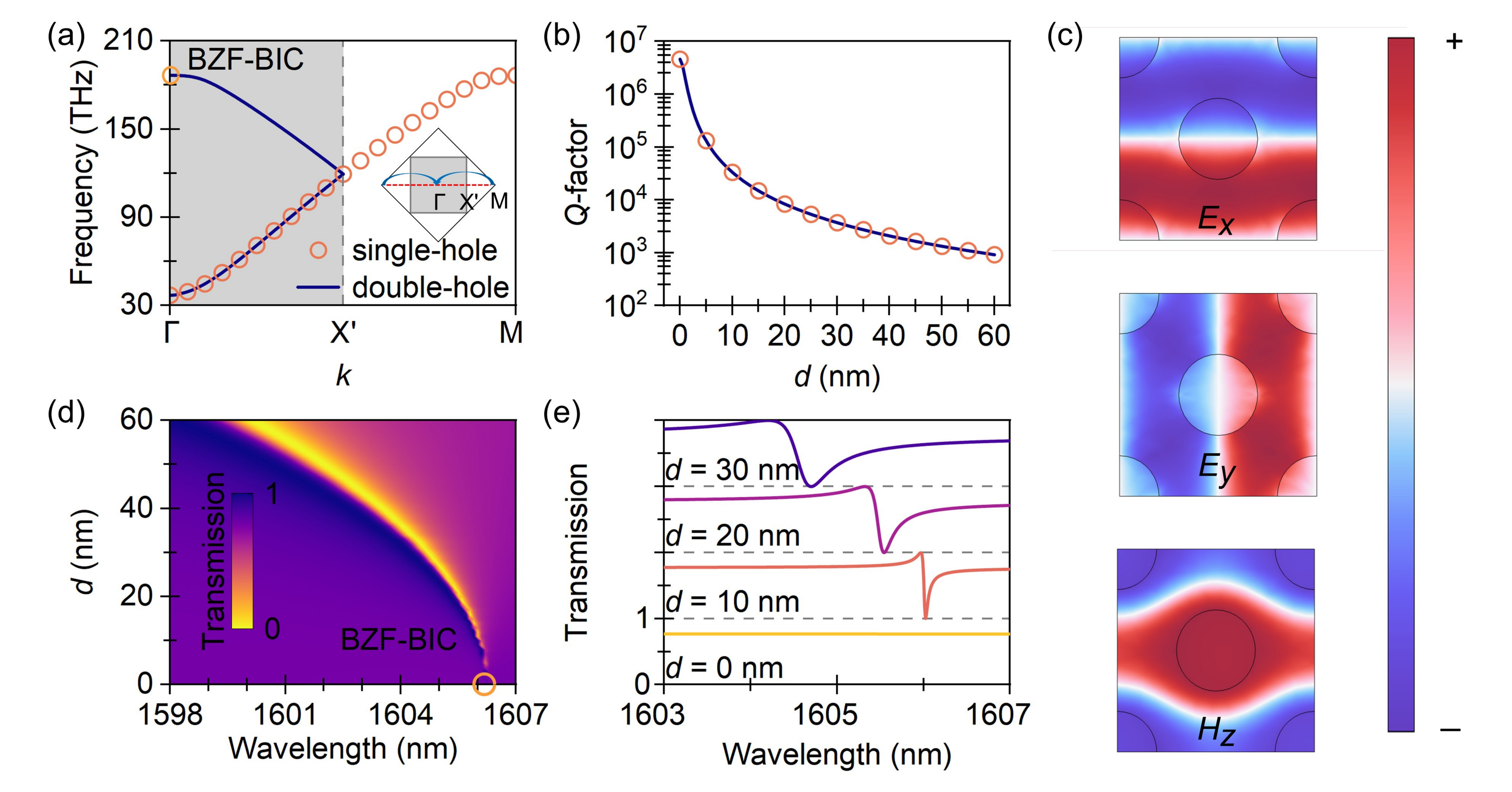}  
	\caption{(a) Band folding schematic of the BZF-BIC from the $\textbf{M}$ point to the $\Gamma$ point (inset: Brillouin zone folding process). (b) $Q$-factor of the BZF-BIC at the $\Gamma$ point versus the perturbation displacement $d$. (c) Electromagnetic field distributions within the metasurface unit cell under Brillouin zone folding; (d) Transmission spectra of the metasurface as functions of wavelength and the perturbation $d$; (e) Transmission spectra for selected $d$ values.}
	\label{Fig2}
\end{figure*}

The inset of Fig. \hyperref[Fig2]{2}(a) schematically illustrates the BZF process induced by periodic structural perturbations. By displacing the central air hole along the $y$-axis, the $\textbf{M}$ point at the Brillouin zone boundary folds along the $\textbf{M}$-$\textbf{X}$'-$\Gamma$ path toward the $\Gamma$ point, generating a new type of BIC termed BZF-BIC. To validate this mechanism, we perform finite element simulations in COMSOL Multiphysics with periodic boundary conditions along the $x$ and $y$ directions (modeling infinite periodicity), and perfectly matched layers (PMLs) in the $z$ direction (see Supplemental Material Section 3). Noted, LN’s refractive index is modeled with full dispersion consideration rather than as a constant, essential for capturing electro-optic modulation effects in later sections (see Supplemental Material Section 1). As shown in Fig. \hyperref[Fig2]{2}(a), the single-hole structure exhibits a guided mode at the $\textbf{M}$ point with a eigenfrequency of $\sim186$ THz. Upon introducing perturbations (double-hole configuration), this mode’s energy band folds through $\textbf{X}$' to $\Gamma$, which is consistent with the BZF process—confirming BZF-BIC formation in the inset (see Supplemental Material Section 2). Figure \hyperref[Fig2]{2}(b) demonstrates that the mode’s $Q$-factor follows the inverse square law relationship, aligning with symmetry-protected BIC scaling laws (see Supplemental Material Section 4)\cite{Overvig2020}. Figure \hyperref[Fig2]{2}(c) further reveals the mode’s electromagnetic field symmetry at the $\Gamma$ point: the electric field components $E_{x}$ and $E_{y}$ exhibit odd symmetry, while the magnetic field component $H_{z}$ shows even symmetry. These features provide conclusive evidence for the formation of the BZF-BIC.  

Building on the eigenfrequency analysis above, we investigate the linear spectral response of this BZF-BIC in the metasurface (see Supplemental Material Section 5). The transmission spectra are simulated as a function of the periodic perturbation displacement $d$, as shown in Fig. \hyperref[Fig2]{2}(d). Increasing $d$ causes the resonance linewidth to broaden from zero, indicating a transition from the non-radiative BIC to a radiative quasi-BIC with a finite $Q$-factor. Specifically, at $d=0$, the BIC exhibits no radiation loss (theoretical linewidth = 0), while perturbations at $d>0$ partially break its confinement, enabling energy leakage into free space and forming a quasi-BIC. Figure \hyperref[Fig2]{2}(e) further reveals the impact of $d$ on transmission properties and resonance wavelength. At $d=0$, no resonance occurs within the wavelength range of interest, consistent with the BIC’s non-radiative nature. As $d$ increases, resonance peaks emerge with blue shifts and progressively broader linewidths, demonstrating enhanced radiative losses and reduced $Q$-factors at larger $d$. Theoretically, smaller $d$ values yield higher $Q$-factors and narrower linewidths, which are critical for optimizing SHG conversion efficiency and electro-optic modulation depth.

\subsection{Nonlinear SHG}

Next, we proceed to discuss the SHG in the LN metasurface. Its second-order nonlinear polarization is expressed as\cite{Tu2024} 

\begin{equation}
\vec{P}^{(2\omega)} = \varepsilon_{0}\chi^{(2)}\vec{E}^{(\omega)}\vec{E}^{(\omega)},
\end{equation}
where $\varepsilon_{0}$ is the vacuum permittivity, and $\chi^{(2)}$ is LN's second-order nonlinear susceptibility tensor. For $x$-cut LN thin film, the contracted matrix form of $\chi^{(2)}$ under Kleinman symmetry is\cite{He2024a}

\begin{equation}
\chi^{(2)} = 2\begin{pmatrix}
d_{22} & 0 & -d_{22} & 0 & 0 & d_{31} \\
d_{31} & d_{33} & d_{31} & 0 & 0 & 0 \\
0 & 0 & 0 & d_{31} & -d_{22} &0
\end{pmatrix},
\end{equation}
with coefficients $d_{22}=2.1$ pm/V, $d_{31}=4.1$ pm/V, and $d_{33}=22.8$ pm/V\cite{Seres2001}. The SHG conversion efficiency is defined as

\begin{equation}
\eta = \frac{P_{\text{out}}}{P_{\text{in}}},
\end{equation}
where $P_{\text{out}}$ and $P_{\text{in}}$ denote the SHG output power and input pump power, respectively.

\begin{figure}[t]
	\centering
	\includegraphics[width=0.5\textwidth]{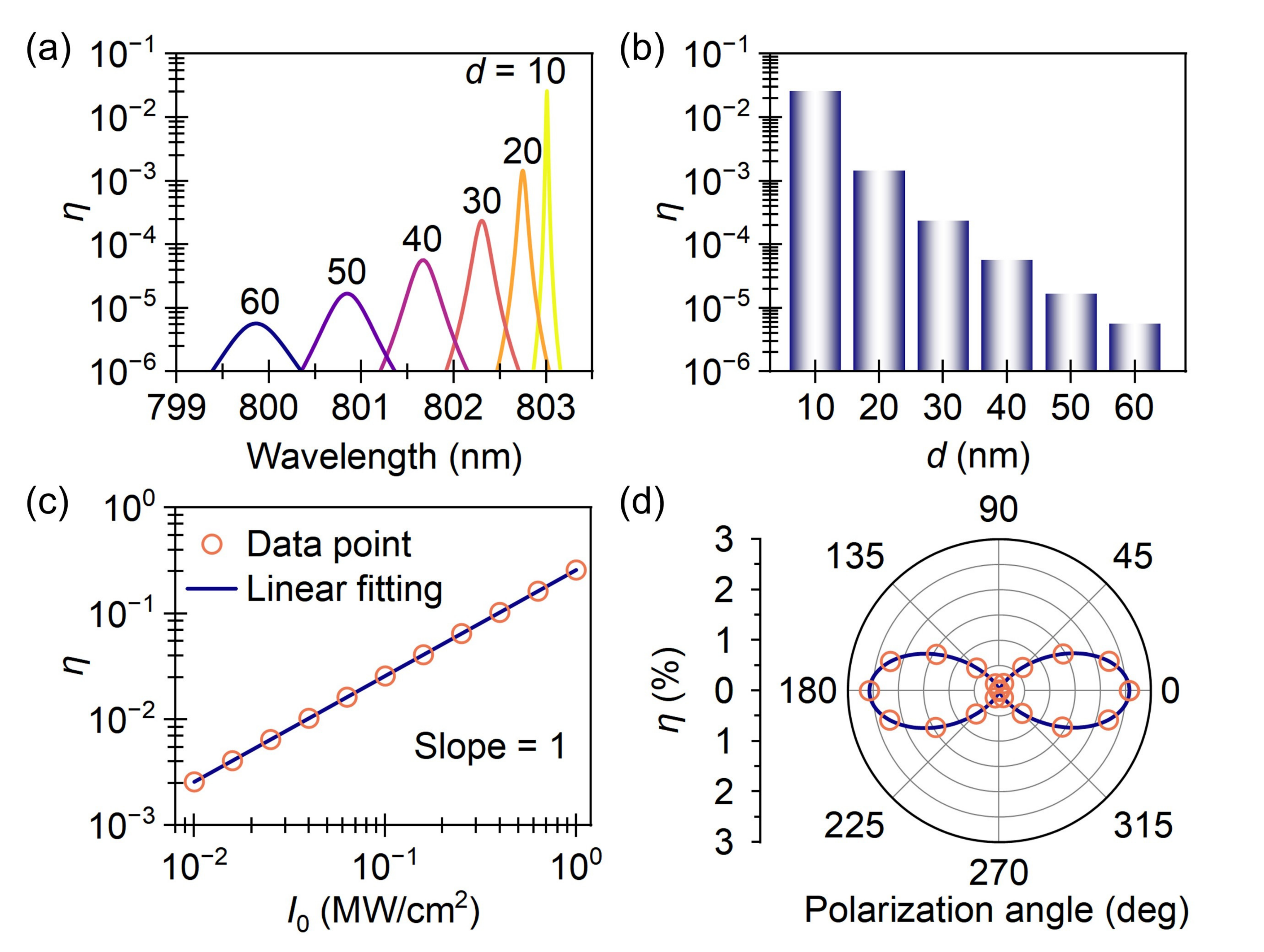}  
	\caption{(a) SHG spectra of the metasurface under an input pump intensity of 0.1 MW/cm$^{2}$ for perturbation displacements ranging from $d=10$ to 60 nm. (b) SHG conversion efficiency $\eta$ for selected $d$ values. (c) Scaling dependence of $\eta$ on the input pump intensity $I_{0}$. (d) Polarization dependence of $\eta$ on the $x$- and $y$-polarized pumps.}
	\label{Fig3}
\end{figure}

We calculate the SHG spectra of the metasurface under a pump intensity of 0.1 MW/cm$^{2}$ for different perturbation displacements. As shown in Fig. \hyperref[Fig3]{3}(a), $\eta$ increases as $d$ decreases. This trend is closely related to the geometric asymmetry, the $Q$-factor of the BZF-BIC, and the enhanced field confinement at smaller $d$. Figure \hyperref[Fig3]{3}(b) explicitly correlates $\eta$ with $d$. Theoretically, as $d$ approaches zero, the $Q$-factor tends toward infinity, significantly enhancing the local field and enabling $\eta$ to reach extremely high levels. However, in practical fabrication processes, unavoidable manufacturing errors can degrade metasurface performance, necessitating a trade-off between achieving a high $Q$-factor and ensuring fabrication feasibility. When $d = 10$ nm (a value achievable under current fabrication conditions\cite{Wang2015, Li2019}), the metasurface achieves an SHG efficiency of $2.59\%$.

On the other hand, $\eta$ is also influenced by the pump intensity. As shown in Fig. \hyperref[Fig3]{3}(c), $\eta$ exhibits a linear relationship with the pump intensity $I_{0}$, indicating that increasing $I_{0}$ can enhance $\eta$ by several orders of magnitude. This result aligns well with previous theoretical predictions of second-order nonlinear process enhancement based on quasi-BIC\cite{Liu2019, Liu2023, Feng2023, Tu2024, He2024a}. However, if $I_{0}$ continues to increase, $\eta$ will be affected by backward frequency conversion and thus no longer exhibits linear enhancement (see Supplemental Material Section 6). Furthermore, we investigated the effect of the pump polarization on $\eta$. As shown in Fig. \hyperref[Fig3]{3}(d), $\eta$ remains high for $x$-polarized pump light but nearly vanishes for $y$-polarized light, demonstrating strong polarization dependence. This phenomenon is primarily due to two factors: First, introducing the perturbation factor $d$ breaks the $C_{4v}$ symmetry of the structure, resulting in differences in linear transmission spectra for different polarization directions and consequently leading to polarization-dependent SHG. Second, in addition to the slight difference in the refrative index in the $x$-$y$ plane, the nonlinear polarization terms are affected differently by $E_x$ and $E_y$. This also contributes to polarization-dependent SHG characteristics.

\subsection{Electro-optic modulation}

Based on the effects of the perturbation displacement $d$, the pump intensity $I_{0}$, and the polarization dependence on the nonlinear conversion efficiency of the LN metasurface discussed in the previous sections, in the following electro-optic modulation, we conduct simulations where $x$-polarized light with a pump intensity of 0.1 MW/cm$^{2}$ is incident on the LN metasurface with $d=10$ nm. Due to the linear Pockels effect in LN, the refractive index as a function of the applied electric field is given by\cite{Gao2021}

\begin{equation}
\left\{
\begin{aligned}
n_{x} = n_{z} = n_{\text{o}} - 
\frac{1}{2}r_{13}n_{\text{o}}^{3}E_{\text{EO}},\\
n_{y} = n_{\text{e}} - 
\frac{1}{2}r_{33}n_{\text{e}}^{3}E_{\text{EO}},\\
\end{aligned}
\right.
\end{equation}
where $r_{13}$ and $r_{33}$ are LN's electro-optic coefficients with values of 10.3 pm/V and 34.1 pm/V, respectively\cite{DiFrancescantonio2024}, and $E_{\text{EO}}$ represents the applied electric field defined with the modulation voltage $V_{\text{pp}}$ and the distance between the two Au electrodes $L$ as $E_{\text{EO}}=V_{\text{pp}}/L$. In our setup, $L$ is assumed to be 10 $\mu$m.

\begin{figure}[t]
	\centering
	\includegraphics[width=0.5\textwidth]{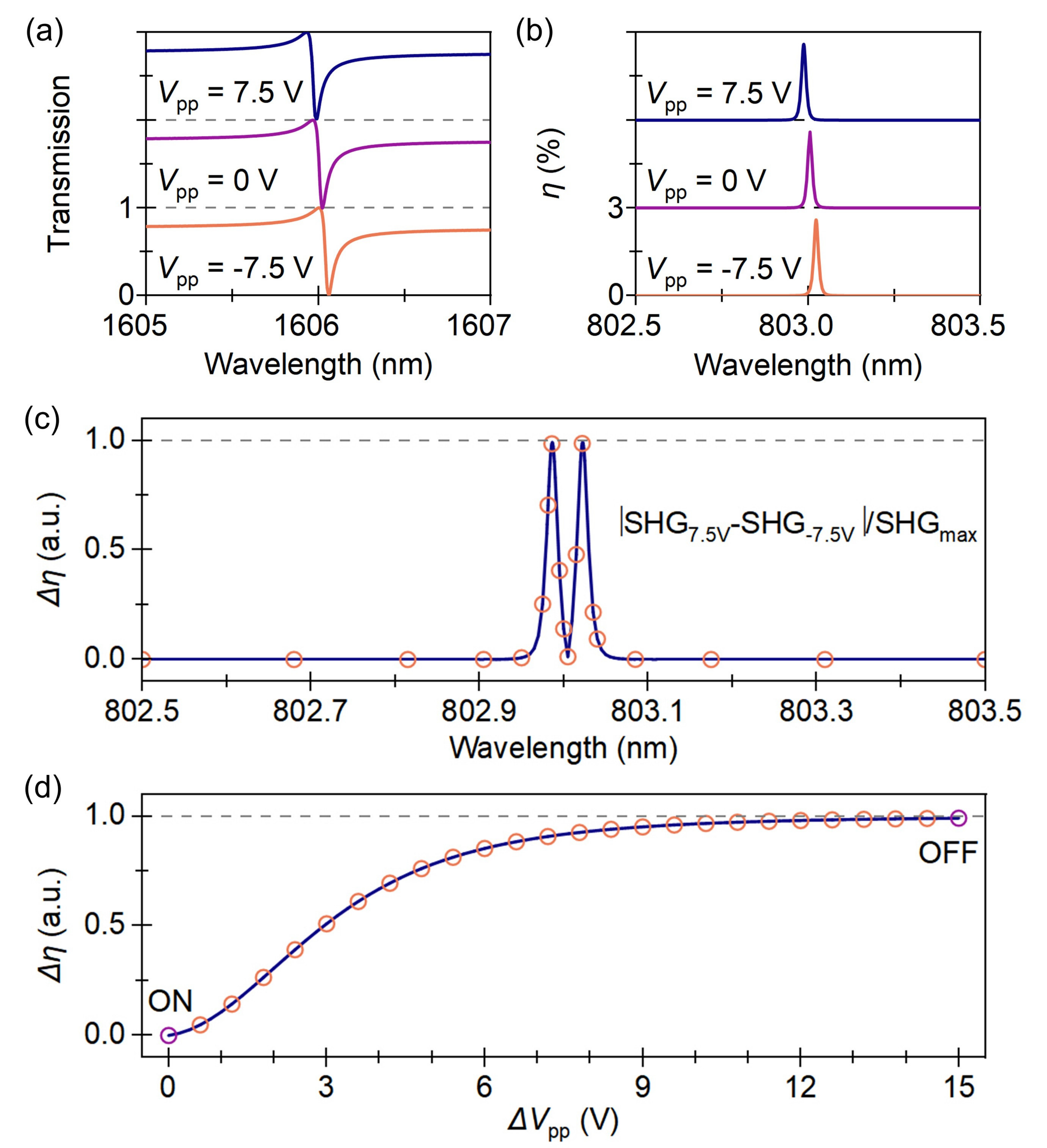}  
	\caption{(a) Transmission and (b) SHG spectra of the metasurface for selected the modulation voltage $V_{\text{pp}}$. (c) Normalized SHG efficiency difference at different wavelengths under $\pm7.5$ V modulation voltage. (d) Normalized SHG efficiency difference under different modulation voltage differences.}
	\label{Fig4}
\end{figure}

Figure \hyperref[Fig4]{4}(a) clearly shows that applying a positive voltage induces a blue shift in the resonance wavelength, while a negative voltage causes a red shift. This behavior stems from the negative linear correlation between LN's refractive index and the applied voltage. The resonance wavelength shift directly affects the spectral position of the SHG, as demonstrated in Fig. \hyperref[Fig4]{4}(b), which displays SHG spectral shifts under $\pm7.5$ V applied voltage. To visualize the switching modulation effect of the LN metasurface at $\pm7.5$ V, Fig. \hyperref[Fig4]{4}(c) plots the absolute difference between the SHG efficiency $\eta$ at $-7.5$ V and $+7.5$ V, revealing an '$\textbf{M}$'-shaped distribution. After normalizing the data for quantitative analysis, the modulation depth approaches unity, confirming that the on-to-off SHG switching control is achieved with a 15 V voltage difference.

\begin{table*}[t]
	\caption{\label{tab:table1}%
		Comparison of electro-optic SHG modulation in LN metasurfaces}
	\begin{ruledtabular}
		\begin{tabular}{ccccc}
	                   & Resonance type  & Resonance wavelength (nm) & Modulation voltage (V) & On-to-off Modulation \\
			\hline
			\cite{He2024}  & Guided mode              & 902                       & 100                    & NO \\
			\cite{DiFrancescantonio2024}  & symmetry-protected BIC             & 1554                      & 18                     & NO \\
			\cite{Kanyang2025}  & Super Fano      & 1308                      & 16                     & YES \\
			This work  & BZF-BIC         & 1606                      & 15                     & YES \\
		\end{tabular}
	\end{ruledtabular}
\end{table*}

We further examine the LN metasurface's modulation performance under varying applied voltages. Figure \hyperref[Fig4]{4}(d) plots the modulation depth versus voltage difference in the range of 0–15 V, showing a progressive enhancement of the modulation effect with increasing voltage until saturation occurs at $\Delta V_{\text{pp}}=15$ V, where stable switching modulation is attained. Further voltage increases yield negligible improvement. Additionally, the modulation voltage cannot exceed LN's damage threshold. Within this material-specific limit, electric-field-induced resonance wavelength shifts enable precise tuning of SHG peaks to target wavelengths. Compared to structural tuning via perturbation modifications, this electro-optic approach avoids the cumbersome re-fabrication process while preserving high SHG efficiency, fully demonstrating reconfigurable metasurfaces' capability for nonlinear optical control.

Finally, to highlight the advantages of this work, Table 1 compares our results with several relevant studies. By evaluating both the required modulation voltages and the achievement of SHG switching modulation, we demonstrate that our work realizes the on-to-off SHG modulation at significantly lower voltages than previous approaches. This performance underscores the superior capabilities of high-$Q$ BZF-BIC for SHG control applications.

\section{Conclusions}

In conclusion, we have excited the BZF-BIC in the LN metasurface for dynamical nonlinear applications. By leveraging its ultrahigh $Q$-factor and narrow spectral linewidth, we successfully achieve high SHG conversion efficiency and on-to-off SHG modulation at a voltage of $\Delta V_{\text{pp}}=15$ V. Although our results are obtained under ideal conditions using finite element simulations, we thoroughly account for the factors that could influence the results in real-world scenarios during the simulation process, ensuring that our conclusions remain feasible and reliable. It is also worth pointing out that the design and mechanism in this work are general, and can be extended to other dynamical nonlinear optical effects, in particular the photon pair generation via quantum frequency conversion\cite{Zhang2022, Ma2023, Liu2024, Ma2024}. From this perspective, our work offers critical insights into the development of on-chip applications in high-performance nonlinear and quantum optical devices. 

\begin{acknowledgments}
	
This work was supported by the National Natural Science Foundation of China (Grants No. 12304420, No. 12264028, No. 12364045, No. 12364049, and No. 12104105), the Natural Science Foundation of Jiangxi Province (Grants No. 20232BAB201040, No. 20232BAB211025, and No. 20242BAB25041), and the Young Elite Scientists Sponsorship Program by JXAST (Grants No. 2023QT11 and No. 2025QT04).

\end{acknowledgments}

\nocite{*}


\begin{thebibliography}{68}%
	\makeatletter
	\providecommand \@ifxundefined [1]{%
		\@ifx{#1\undefined}
	}%
	\providecommand \@ifnum [1]{%
		\ifnum #1\expandafter \@firstoftwo
		\else \expandafter \@secondoftwo
		\fi
	}%
	\providecommand \@ifx [1]{%
		\ifx #1\expandafter \@firstoftwo
		\else \expandafter \@secondoftwo
		\fi
	}%
	\providecommand \natexlab [1]{#1}%
	\providecommand \enquote  [1]{``#1''}%
	\providecommand \bibnamefont  [1]{#1}%
	\providecommand \bibfnamefont [1]{#1}%
	\providecommand \citenamefont [1]{#1}%
	\providecommand \href@noop [0]{\@secondoftwo}%
	\providecommand \href [0]{\begingroup \@sanitize@url \@href}%
	\providecommand \@href[1]{\@@startlink{#1}\@@href}%
	\providecommand \@@href[1]{\endgroup#1\@@endlink}%
	\providecommand \@sanitize@url [0]{\catcode `\\12\catcode `\$12\catcode
		`\&12\catcode `\#12\catcode `\^12\catcode `\_12\catcode `\%12\relax}%
	\providecommand \@@startlink[1]{}%
	\providecommand \@@endlink[0]{}%
	\providecommand \url  [0]{\begingroup\@sanitize@url \@url }%
	\providecommand \@url [1]{\endgroup\@href {#1}{\urlprefix }}%
	\providecommand \urlprefix  [0]{URL }%
	\providecommand \Eprint [0]{\href }%
	\providecommand \doibase [0]{https://doi.org/}%
	\providecommand \selectlanguage [0]{\@gobble}%
	\providecommand \bibinfo  [0]{\@secondoftwo}%
	\providecommand \bibfield  [0]{\@secondoftwo}%
	\providecommand \translation [1]{[#1]}%
	\providecommand \BibitemOpen [0]{}%
	\providecommand \bibitemStop [0]{}%
	\providecommand \bibitemNoStop [0]{.\EOS\space}%
	\providecommand \EOS [0]{\spacefactor3000\relax}%
	\providecommand \BibitemShut  [1]{\csname bibitem#1\endcsname}%
	\let\auto@bib@innerbib\@empty
	\bibitem [{\citenamefont {Kildishev}\ \emph {et~al.}(2013)\citenamefont
		{Kildishev}, \citenamefont {Boltasseva},\ and\ \citenamefont
		{Shalaev}}]{Kildishev2013}%
	\BibitemOpen
	\bibfield  {author} {\bibinfo {author} {\bibfnamefont {A.~V.}\ \bibnamefont
			{Kildishev}}, \bibinfo {author} {\bibfnamefont {A.}~\bibnamefont
			{Boltasseva}},\ and\ \bibinfo {author} {\bibfnamefont {V.~M.}\ \bibnamefont
			{Shalaev}},\ }\bibfield  {title} {\bibinfo {title} {Planar photonics with
			metasurfaces},\ }\href {https://doi.org/10.1126/science.1232009} {\bibfield
		{journal} {\bibinfo  {journal} {Science}\ }\textbf {\bibinfo {volume}
			{339}},\ \bibinfo {pages} {1232009} (\bibinfo {year} {2013})}\BibitemShut
	{NoStop}%
	\bibitem [{\citenamefont {Yu}\ and\ \citenamefont {Capasso}(2014)}]{Yu2014}%
	\BibitemOpen
	\bibfield  {author} {\bibinfo {author} {\bibfnamefont {N.}~\bibnamefont
			{Yu}}\ and\ \bibinfo {author} {\bibfnamefont {F.}~\bibnamefont {Capasso}},\
	}\bibfield  {title} {\bibinfo {title} {Flat optics with designer
			metasurfaces},\ }\href {https://doi.org/10.1038/nmat3839} {\bibfield
		{journal} {\bibinfo  {journal} {Nat. Mater.}\ }\textbf {\bibinfo {volume}
			{13}},\ \bibinfo {pages} {139} (\bibinfo {year} {2014})}\BibitemShut
	{NoStop}%
	\bibitem [{\citenamefont {So}\ \emph {et~al.}(2023)\citenamefont {So},
		\citenamefont {Mun}, \citenamefont {Park},\ and\ \citenamefont
		{Rho}}]{So2023}%
	\BibitemOpen
	\bibfield  {author} {\bibinfo {author} {\bibfnamefont {S.}~\bibnamefont
			{So}}, \bibinfo {author} {\bibfnamefont {J.}~\bibnamefont {Mun}}, \bibinfo
		{author} {\bibfnamefont {J.}~\bibnamefont {Park}},\ and\ \bibinfo {author}
		{\bibfnamefont {J.}~\bibnamefont {Rho}},\ }\bibfield  {title} {\bibinfo
		{title} {Revisiting the design strategies for metasurfaces: Fundamental
			physics, optimization, and beyond},\ }\href
	{https://doi.org/10.1002/adma.202206399} {\bibfield  {journal} {\bibinfo
			{journal} {Adv. Mater.}\ }\textbf {\bibinfo {volume} {35}},\ \bibinfo {pages}
		{2206399} (\bibinfo {year} {2023})}\BibitemShut {NoStop}%
	\bibitem [{\citenamefont {Yao}\ \emph {et~al.}(2023)\citenamefont {Yao},
		\citenamefont {Lin}, \citenamefont {Chen},\ and\ \citenamefont
		{Tsai}}]{Yao2023}%
	\BibitemOpen
	\bibfield  {author} {\bibinfo {author} {\bibfnamefont {J.}~\bibnamefont
			{Yao}}, \bibinfo {author} {\bibfnamefont {R.}~\bibnamefont {Lin}}, \bibinfo
		{author} {\bibfnamefont {M.~K.}\ \bibnamefont {Chen}},\ and\ \bibinfo
		{author} {\bibfnamefont {D.~P.}\ \bibnamefont {Tsai}},\ }\bibfield  {title}
	{\bibinfo {title} {Integrated-resonant metadevices: a review},\ }\href
	{https://doi.org/10.1117/1.ap.5.2.024001} {\bibfield  {journal} {\bibinfo
			{journal} {Adv. Photonics}\ }\textbf {\bibinfo {volume} {5}},\ \bibinfo
		{pages} {024001} (\bibinfo {year} {2023})}\BibitemShut {NoStop}%
	\bibitem [{\citenamefont {Kuznetsov}\ \emph {et~al.}(2024)\citenamefont
		{Kuznetsov}, \citenamefont {Brongersma}, \citenamefont {Yao}, \citenamefont
		{Chen}, \citenamefont {Levy}, \citenamefont {Tsai}, \citenamefont {Zheludev},
		\citenamefont {Faraon}, \citenamefont {Arbabi}, \citenamefont {Yu},
		\citenamefont {Chanda}, \citenamefont {Crozier}, \citenamefont {Kildishev},
		\citenamefont {Wang}, \citenamefont {Yang}, \citenamefont {Valentine},
		\citenamefont {Genevet}, \citenamefont {Fan}, \citenamefont {Miller},
		\citenamefont {Majumdar}, \citenamefont {Fröch}, \citenamefont {Brady},
		\citenamefont {Heide}, \citenamefont {Veeraraghavan}, \citenamefont
		{Engheta}, \citenamefont {Alù}, \citenamefont {Polman}, \citenamefont
		{Atwater}, \citenamefont {Thureja}, \citenamefont {Paniagua-Dominguez},
		\citenamefont {Ha}, \citenamefont {Barreda}, \citenamefont {Schuller},
		\citenamefont {Staude}, \citenamefont {Grinblat}, \citenamefont {Kivshar},
		\citenamefont {Peana}, \citenamefont {Yelin}, \citenamefont {Senichev},
		\citenamefont {Shalaev}, \citenamefont {Saha}, \citenamefont {Boltasseva},
		\citenamefont {Rho}, \citenamefont {Oh}, \citenamefont {Kim}, \citenamefont
		{Park}, \citenamefont {Devlin},\ and\ \citenamefont {Pala}}]{Kuznetsov2024}%
	\BibitemOpen
	\bibfield  {author} {\bibinfo {author} {\bibfnamefont {A.~I.}\ \bibnamefont
			{Kuznetsov}}, \bibinfo {author} {\bibfnamefont {M.~L.}\ \bibnamefont
			{Brongersma}}, \bibinfo {author} {\bibfnamefont {J.}~\bibnamefont {Yao}},
		\bibinfo {author} {\bibfnamefont {M.~K.}\ \bibnamefont {Chen}}, \bibinfo
		{author} {\bibfnamefont {U.}~\bibnamefont {Levy}}, \bibinfo {author}
		{\bibfnamefont {D.~P.}\ \bibnamefont {Tsai}}, \bibinfo {author}
		{\bibfnamefont {N.~I.}\ \bibnamefont {Zheludev}}, \bibinfo {author}
		{\bibfnamefont {A.}~\bibnamefont {Faraon}}, \bibinfo {author} {\bibfnamefont
			{A.}~\bibnamefont {Arbabi}}, \bibinfo {author} {\bibfnamefont
			{N.}~\bibnamefont {Yu}}, \bibinfo {author} {\bibfnamefont {D.}~\bibnamefont
			{Chanda}}, \bibinfo {author} {\bibfnamefont {K.~B.}\ \bibnamefont {Crozier}},
		\bibinfo {author} {\bibfnamefont {A.~V.}\ \bibnamefont {Kildishev}}, \bibinfo
		{author} {\bibfnamefont {H.}~\bibnamefont {Wang}}, \bibinfo {author}
		{\bibfnamefont {J.~K.~W.}\ \bibnamefont {Yang}}, \bibinfo {author}
		{\bibfnamefont {J.~G.}\ \bibnamefont {Valentine}}, \bibinfo {author}
		{\bibfnamefont {P.}~\bibnamefont {Genevet}}, \bibinfo {author} {\bibfnamefont
			{J.~A.}\ \bibnamefont {Fan}}, \bibinfo {author} {\bibfnamefont {O.~D.}\
			\bibnamefont {Miller}}, \bibinfo {author} {\bibfnamefont {A.}~\bibnamefont
			{Majumdar}}, \bibinfo {author} {\bibfnamefont {J.~E.}\ \bibnamefont
			{Fröch}}, \bibinfo {author} {\bibfnamefont {D.}~\bibnamefont {Brady}},
		\bibinfo {author} {\bibfnamefont {F.}~\bibnamefont {Heide}}, \bibinfo
		{author} {\bibfnamefont {A.}~\bibnamefont {Veeraraghavan}}, \bibinfo {author}
		{\bibfnamefont {N.}~\bibnamefont {Engheta}}, \bibinfo {author} {\bibfnamefont
			{A.}~\bibnamefont {Alù}}, \bibinfo {author} {\bibfnamefont {A.}~\bibnamefont
			{Polman}}, \bibinfo {author} {\bibfnamefont {H.~A.}\ \bibnamefont {Atwater}},
		\bibinfo {author} {\bibfnamefont {P.}~\bibnamefont {Thureja}}, \bibinfo
		{author} {\bibfnamefont {R.}~\bibnamefont {Paniagua-Dominguez}}, \bibinfo
		{author} {\bibfnamefont {S.~T.}\ \bibnamefont {Ha}}, \bibinfo {author}
		{\bibfnamefont {A.~I.}\ \bibnamefont {Barreda}}, \bibinfo {author}
		{\bibfnamefont {J.~A.}\ \bibnamefont {Schuller}}, \bibinfo {author}
		{\bibfnamefont {I.}~\bibnamefont {Staude}}, \bibinfo {author} {\bibfnamefont
			{G.}~\bibnamefont {Grinblat}}, \bibinfo {author} {\bibfnamefont
			{Y.}~\bibnamefont {Kivshar}}, \bibinfo {author} {\bibfnamefont
			{S.}~\bibnamefont {Peana}}, \bibinfo {author} {\bibfnamefont {S.~F.}\
			\bibnamefont {Yelin}}, \bibinfo {author} {\bibfnamefont {A.}~\bibnamefont
			{Senichev}}, \bibinfo {author} {\bibfnamefont {V.~M.}\ \bibnamefont
			{Shalaev}}, \bibinfo {author} {\bibfnamefont {S.}~\bibnamefont {Saha}},
		\bibinfo {author} {\bibfnamefont {A.}~\bibnamefont {Boltasseva}}, \bibinfo
		{author} {\bibfnamefont {J.}~\bibnamefont {Rho}}, \bibinfo {author}
		{\bibfnamefont {D.~K.}\ \bibnamefont {Oh}}, \bibinfo {author} {\bibfnamefont
			{J.}~\bibnamefont {Kim}}, \bibinfo {author} {\bibfnamefont {J.}~\bibnamefont
			{Park}}, \bibinfo {author} {\bibfnamefont {R.}~\bibnamefont {Devlin}},\ and\
		\bibinfo {author} {\bibfnamefont {R.~A.}\ \bibnamefont {Pala}},\ }\bibfield
	{title} {\bibinfo {title} {Roadmap for optical metasurfaces},\ }\href
	{https://doi.org/10.1021/acsphotonics.3c00457} {\bibfield  {journal}
		{\bibinfo  {journal} {ACS Photonics}\ }\textbf {\bibinfo {volume} {11}},\
		\bibinfo {pages} {816} (\bibinfo {year} {2024})}\BibitemShut {NoStop}%
	\bibitem [{\citenamefont {Hu}\ \emph {et~al.}(2023)\citenamefont {Hu},
		\citenamefont {Jiang}, \citenamefont {Zhang}, \citenamefont {Yang},
		\citenamefont {Ou}, \citenamefont {Li}, \citenamefont {Kong}, \citenamefont
		{Liu}, \citenamefont {Qiu},\ and\ \citenamefont {Duan}}]{Hu2023}%
	\BibitemOpen
	\bibfield  {author} {\bibinfo {author} {\bibfnamefont {Y.}~\bibnamefont
			{Hu}}, \bibinfo {author} {\bibfnamefont {Y.}~\bibnamefont {Jiang}}, \bibinfo
		{author} {\bibfnamefont {Y.}~\bibnamefont {Zhang}}, \bibinfo {author}
		{\bibfnamefont {X.}~\bibnamefont {Yang}}, \bibinfo {author} {\bibfnamefont
			{X.}~\bibnamefont {Ou}}, \bibinfo {author} {\bibfnamefont {L.}~\bibnamefont
			{Li}}, \bibinfo {author} {\bibfnamefont {X.}~\bibnamefont {Kong}}, \bibinfo
		{author} {\bibfnamefont {X.}~\bibnamefont {Liu}}, \bibinfo {author}
		{\bibfnamefont {C.-W.}\ \bibnamefont {Qiu}},\ and\ \bibinfo {author}
		{\bibfnamefont {H.}~\bibnamefont {Duan}},\ }\bibfield  {title} {\bibinfo
		{title} {Asymptotic dispersion engineering for ultra-broadband meta-optics},\
	}\href {https://doi.org/10.1038/s41467-023-42268-5} {\bibfield  {journal}
		{\bibinfo  {journal} {Nat. Commun.}\ }\textbf {\bibinfo {volume} {14}},\
		\bibinfo {pages} {6649} (\bibinfo {year} {2023})}\BibitemShut {NoStop}%
	\bibitem [{\citenamefont {Hou}\ \emph {et~al.}(2025)\citenamefont {Hou},
		\citenamefont {Zhou}, \citenamefont {Zhang}, \citenamefont {Lu},
		\citenamefont {Zhang}, \citenamefont {Liu}, \citenamefont {Xiao},\ and\
		\citenamefont {Yu}}]{Hou2025}%
	\BibitemOpen
	\bibfield  {author} {\bibinfo {author} {\bibfnamefont {L.}~\bibnamefont
			{Hou}}, \bibinfo {author} {\bibfnamefont {H.}~\bibnamefont {Zhou}}, \bibinfo
		{author} {\bibfnamefont {D.}~\bibnamefont {Zhang}}, \bibinfo {author}
		{\bibfnamefont {G.}~\bibnamefont {Lu}}, \bibinfo {author} {\bibfnamefont
			{D.}~\bibnamefont {Zhang}}, \bibinfo {author} {\bibfnamefont
			{T.}~\bibnamefont {Liu}}, \bibinfo {author} {\bibfnamefont {S.}~\bibnamefont
			{Xiao}},\ and\ \bibinfo {author} {\bibfnamefont {T.}~\bibnamefont {Yu}},\
	}\bibfield  {title} {\bibinfo {title} {High-efficiency broadband achromatic
			metalens in the visible},\ }\href {https://doi.org/10.1063/5.0240728}
	{\bibfield  {journal} {\bibinfo  {journal} {Appl. Phys. Lett.}\ }\textbf
		{\bibinfo {volume} {126}},\ \bibinfo {pages} {101704} (\bibinfo {year}
		{2025})}\BibitemShut {NoStop}%
	\bibitem [{\citenamefont {Li}\ \emph {et~al.}(2020)\citenamefont {Li},
		\citenamefont {Chen}, \citenamefont {Guan}, \citenamefont {Tao},
		\citenamefont {Chang}, \citenamefont {Dai}, \citenamefont {Xiao},
		\citenamefont {Cui}, \citenamefont {Wang}, \citenamefont {Yu}, \citenamefont
		{Zheng},\ and\ \citenamefont {Zhang}}]{Li2020}%
	\BibitemOpen
	\bibfield  {author} {\bibinfo {author} {\bibfnamefont {Z.}~\bibnamefont
			{Li}}, \bibinfo {author} {\bibfnamefont {C.}~\bibnamefont {Chen}}, \bibinfo
		{author} {\bibfnamefont {Z.}~\bibnamefont {Guan}}, \bibinfo {author}
		{\bibfnamefont {J.}~\bibnamefont {Tao}}, \bibinfo {author} {\bibfnamefont
			{S.}~\bibnamefont {Chang}}, \bibinfo {author} {\bibfnamefont
			{Q.}~\bibnamefont {Dai}}, \bibinfo {author} {\bibfnamefont {Y.}~\bibnamefont
			{Xiao}}, \bibinfo {author} {\bibfnamefont {Y.}~\bibnamefont {Cui}}, \bibinfo
		{author} {\bibfnamefont {Y.}~\bibnamefont {Wang}}, \bibinfo {author}
		{\bibfnamefont {S.}~\bibnamefont {Yu}}, \bibinfo {author} {\bibfnamefont
			{G.}~\bibnamefont {Zheng}},\ and\ \bibinfo {author} {\bibfnamefont
			{S.}~\bibnamefont {Zhang}},\ }\bibfield  {title} {\bibinfo {title}
		{Three‐channel metasurfaces for simultaneous meta‐holography and
			meta‐nanoprinting: A single‐cell design approach},\ }\href
	{https://doi.org/10.1002/lpor.202000032} {\bibfield  {journal} {\bibinfo
			{journal} {Laser Photonics Rev.}\ }\textbf {\bibinfo {volume} {14}},\
		\bibinfo {pages} {2000032} (\bibinfo {year} {2020})}\BibitemShut {NoStop}%
	\bibitem [{\citenamefont {Meng}\ \emph {et~al.}(2025)\citenamefont {Meng},
		\citenamefont {Fröch}, \citenamefont {Cheng}, \citenamefont {Pi},
		\citenamefont {Li}, \citenamefont {Majumdar}, \citenamefont {Maier},
		\citenamefont {Ren}, \citenamefont {Gu},\ and\ \citenamefont
		{Fang}}]{Meng2025}%
	\BibitemOpen
	\bibfield  {author} {\bibinfo {author} {\bibfnamefont {W.}~\bibnamefont
			{Meng}}, \bibinfo {author} {\bibfnamefont {J.~E.}\ \bibnamefont {Fröch}},
		\bibinfo {author} {\bibfnamefont {K.}~\bibnamefont {Cheng}}, \bibinfo
		{author} {\bibfnamefont {D.}~\bibnamefont {Pi}}, \bibinfo {author}
		{\bibfnamefont {B.}~\bibnamefont {Li}}, \bibinfo {author} {\bibfnamefont
			{A.}~\bibnamefont {Majumdar}}, \bibinfo {author} {\bibfnamefont {S.~A.}\
			\bibnamefont {Maier}}, \bibinfo {author} {\bibfnamefont {H.}~\bibnamefont
			{Ren}}, \bibinfo {author} {\bibfnamefont {M.}~\bibnamefont {Gu}},\ and\
		\bibinfo {author} {\bibfnamefont {X.}~\bibnamefont {Fang}},\ }\bibfield
	{title} {\bibinfo {title} {Ultranarrow-linewidth wavelength-vortex
			metasurface holography},\ }\href {https://doi.org/10.1126/sciadv.adt9159}
	{\bibfield  {journal} {\bibinfo  {journal} {Sci. Adv.}\ }\textbf {\bibinfo
			{volume} {11}},\ \bibinfo {pages} {eadt9159} (\bibinfo {year}
		{2025})}\BibitemShut {NoStop}%
	\bibitem [{\citenamefont {Ding}\ \emph {et~al.}(2020)\citenamefont {Ding},
		\citenamefont {Chang}, \citenamefont {Wei}, \citenamefont {Huang},
		\citenamefont {Guan},\ and\ \citenamefont {Bozhevolnyi}}]{Ding2020}%
	\BibitemOpen
	\bibfield  {author} {\bibinfo {author} {\bibfnamefont {F.}~\bibnamefont
			{Ding}}, \bibinfo {author} {\bibfnamefont {B.}~\bibnamefont {Chang}},
		\bibinfo {author} {\bibfnamefont {Q.}~\bibnamefont {Wei}}, \bibinfo {author}
		{\bibfnamefont {L.}~\bibnamefont {Huang}}, \bibinfo {author} {\bibfnamefont
			{X.}~\bibnamefont {Guan}},\ and\ \bibinfo {author} {\bibfnamefont {S.~I.}\
			\bibnamefont {Bozhevolnyi}},\ }\bibfield  {title} {\bibinfo {title}
		{Versatile polarization generation and manipulation using dielectric
			metasurfaces},\ }\href {https://doi.org/10.1002/lpor.202000116} {\bibfield
		{journal} {\bibinfo  {journal} {Laser Photonics Rev.}\ }\textbf {\bibinfo
			{volume} {14}},\ \bibinfo {pages} {2000116} (\bibinfo {year}
		{2020})}\BibitemShut {NoStop}%
	\bibitem [{\citenamefont {Wang}\ \emph {et~al.}(2021)\citenamefont {Wang},
		\citenamefont {Deng}, \citenamefont {Wang}, \citenamefont {Zhou},
		\citenamefont {Wang}, \citenamefont {Cao}, \citenamefont {Guan},
		\citenamefont {Xiao},\ and\ \citenamefont {Li}}]{Wang2021a}%
	\BibitemOpen
	\bibfield  {author} {\bibinfo {author} {\bibfnamefont {S.}~\bibnamefont
			{Wang}}, \bibinfo {author} {\bibfnamefont {Z.-L.}\ \bibnamefont {Deng}},
		\bibinfo {author} {\bibfnamefont {Y.}~\bibnamefont {Wang}}, \bibinfo {author}
		{\bibfnamefont {Q.}~\bibnamefont {Zhou}}, \bibinfo {author} {\bibfnamefont
			{X.}~\bibnamefont {Wang}}, \bibinfo {author} {\bibfnamefont {Y.}~\bibnamefont
			{Cao}}, \bibinfo {author} {\bibfnamefont {B.-O.}\ \bibnamefont {Guan}},
		\bibinfo {author} {\bibfnamefont {S.}~\bibnamefont {Xiao}},\ and\ \bibinfo
		{author} {\bibfnamefont {X.}~\bibnamefont {Li}},\ }\bibfield  {title}
	{\bibinfo {title} {Arbitrary polarization conversion dichroism metasurfaces
			for all-in-one full poincaré sphere polarizers},\ }\href
	{https://doi.org/10.1038/s41377-021-00468-y} {\bibfield  {journal} {\bibinfo
			{journal} {Light Sci. Appl.}\ }\textbf {\bibinfo {volume} {10}},\ \bibinfo
		{pages} {24} (\bibinfo {year} {2021})}\BibitemShut {NoStop}%
	\bibitem [{\citenamefont {Kruk}\ \emph {et~al.}(2022)\citenamefont {Kruk},
		\citenamefont {Wang}, \citenamefont {Sain}, \citenamefont {Dong},
		\citenamefont {Yang}, \citenamefont {Zentgraf},\ and\ \citenamefont
		{Kivshar}}]{Kruk2022}%
	\BibitemOpen
	\bibfield  {author} {\bibinfo {author} {\bibfnamefont {S.~S.}\ \bibnamefont
			{Kruk}}, \bibinfo {author} {\bibfnamefont {L.}~\bibnamefont {Wang}}, \bibinfo
		{author} {\bibfnamefont {B.}~\bibnamefont {Sain}}, \bibinfo {author}
		{\bibfnamefont {Z.}~\bibnamefont {Dong}}, \bibinfo {author} {\bibfnamefont
			{J.}~\bibnamefont {Yang}}, \bibinfo {author} {\bibfnamefont {T.}~\bibnamefont
			{Zentgraf}},\ and\ \bibinfo {author} {\bibfnamefont {Y.}~\bibnamefont
			{Kivshar}},\ }\bibfield  {title} {\bibinfo {title} {Asymmetric parametric
			generation of images with nonlinear dielectric metasurfaces},\ }\href
	{https://doi.org/10.1038/s41566-022-01018-7} {\bibfield  {journal} {\bibinfo
			{journal} {Nat. Photonics}\ }\textbf {\bibinfo {volume} {16}},\ \bibinfo
		{pages} {561} (\bibinfo {year} {2022})}\BibitemShut {NoStop}%
	\bibitem [{\citenamefont {Xiao}\ \emph {et~al.}(2020)\citenamefont {Xiao},
		\citenamefont {Wang}, \citenamefont {Liu}, \citenamefont {Zhou},
		\citenamefont {Jiang},\ and\ \citenamefont {Zhang}}]{Xiao2020}%
	\BibitemOpen
	\bibfield  {author} {\bibinfo {author} {\bibfnamefont {S.}~\bibnamefont
			{Xiao}}, \bibinfo {author} {\bibfnamefont {T.}~\bibnamefont {Wang}}, \bibinfo
		{author} {\bibfnamefont {T.}~\bibnamefont {Liu}}, \bibinfo {author}
		{\bibfnamefont {C.}~\bibnamefont {Zhou}}, \bibinfo {author} {\bibfnamefont
			{X.}~\bibnamefont {Jiang}},\ and\ \bibinfo {author} {\bibfnamefont
			{J.}~\bibnamefont {Zhang}},\ }\bibfield  {title} {\bibinfo {title} {Active
			metamaterials and metadevices: a review},\ }\href
	{https://doi.org/10.1088/1361-6463/abaced} {\bibfield  {journal} {\bibinfo
			{journal} {J. Phys. D: Appl. Phys.}\ }\textbf {\bibinfo {volume} {53}},\
		\bibinfo {pages} {503002} (\bibinfo {year} {2020})}\BibitemShut {NoStop}%
	\bibitem [{\citenamefont {Herle}\ \emph {et~al.}(2023)\citenamefont {Herle},
		\citenamefont {Kiselev}, \citenamefont {Villanueva}, \citenamefont {Martin},\
		and\ \citenamefont {Quack}}]{Herle2023}%
	\BibitemOpen
	\bibfield  {author} {\bibinfo {author} {\bibfnamefont {D.}~\bibnamefont
			{Herle}}, \bibinfo {author} {\bibfnamefont {A.}~\bibnamefont {Kiselev}},
		\bibinfo {author} {\bibfnamefont {L.~G.}\ \bibnamefont {Villanueva}},
		\bibinfo {author} {\bibfnamefont {O.~J.~F.}\ \bibnamefont {Martin}},\ and\
		\bibinfo {author} {\bibfnamefont {N.}~\bibnamefont {Quack}},\ }\bibfield
	{title} {\bibinfo {title} {Broadband mechanically tunable metasurface
			reflectivity modulator in the visible spectrum},\ }\href
	{https://doi.org/10.1021/acsphotonics.3c00276} {\bibfield  {journal}
		{\bibinfo  {journal} {ACS Photonics}\ }\textbf {\bibinfo {volume} {10}},\
		\bibinfo {pages} {1882} (\bibinfo {year} {2023})}\BibitemShut {NoStop}%
	\bibitem [{\citenamefont {Ding}\ \emph {et~al.}(2024)\citenamefont {Ding},
		\citenamefont {Deng}, \citenamefont {Meng}, \citenamefont {Thrane},\ and\
		\citenamefont {Bozhevolnyi}}]{Ding2024}%
	\BibitemOpen
	\bibfield  {author} {\bibinfo {author} {\bibfnamefont {F.}~\bibnamefont
			{Ding}}, \bibinfo {author} {\bibfnamefont {Y.}~\bibnamefont {Deng}}, \bibinfo
		{author} {\bibfnamefont {C.}~\bibnamefont {Meng}}, \bibinfo {author}
		{\bibfnamefont {P.~C.}\ \bibnamefont {Thrane}},\ and\ \bibinfo {author}
		{\bibfnamefont {S.~I.}\ \bibnamefont {Bozhevolnyi}},\ }\bibfield  {title}
	{\bibinfo {title} {Electrically tunable topological phase transition in
			non-hermitian optical mems metasurfaces},\ }\href
	{https://doi.org/10.1126/sciadv.adl4661} {\bibfield  {journal} {\bibinfo
			{journal} {Sci. Adv.}\ }\textbf {\bibinfo {volume} {10}},\ \bibinfo {pages}
		{eadl4661} (\bibinfo {year} {2024})}\BibitemShut {NoStop}%
	\bibitem [{\citenamefont {Eaves-Rathert}\ \emph {et~al.}(2022)\citenamefont
		{Eaves-Rathert}, \citenamefont {Kovalik}, \citenamefont {Ugwu}, \citenamefont
		{Rogers}, \citenamefont {Pint},\ and\ \citenamefont
		{Valentine}}]{EavesRathert2022}%
	\BibitemOpen
	\bibfield  {author} {\bibinfo {author} {\bibfnamefont {J.}~\bibnamefont
			{Eaves-Rathert}}, \bibinfo {author} {\bibfnamefont {E.}~\bibnamefont
			{Kovalik}}, \bibinfo {author} {\bibfnamefont {C.~F.}\ \bibnamefont {Ugwu}},
		\bibinfo {author} {\bibfnamefont {B.~R.}\ \bibnamefont {Rogers}}, \bibinfo
		{author} {\bibfnamefont {C.~L.}\ \bibnamefont {Pint}},\ and\ \bibinfo
		{author} {\bibfnamefont {J.~G.}\ \bibnamefont {Valentine}},\ }\bibfield
	{title} {\bibinfo {title} {Dynamic color tuning with electrochemically
			actuated TiO$_{2}$ metasurfaces},\ }\href
	{https://doi.org/10.1021/acs.nanolett.1c04613} {\bibfield  {journal}
		{\bibinfo  {journal} {Nano Lett.}\ }\textbf {\bibinfo {volume} {22}},\
		\bibinfo {pages} {1626} (\bibinfo {year} {2022})}\BibitemShut {NoStop}%
	\bibitem [{\citenamefont {Kovalik}\ \emph {et~al.}(2024)\citenamefont
		{Kovalik}, \citenamefont {Eaves-Rathert}, \citenamefont {Pint},\ and\
		\citenamefont {Valentine}}]{Kovalik2024}%
	\BibitemOpen
	\bibfield  {author} {\bibinfo {author} {\bibfnamefont {E.}~\bibnamefont
			{Kovalik}}, \bibinfo {author} {\bibfnamefont {J.}~\bibnamefont
			{Eaves-Rathert}}, \bibinfo {author} {\bibfnamefont {C.~L.}\ \bibnamefont
			{Pint}},\ and\ \bibinfo {author} {\bibfnamefont {J.~G.}\ \bibnamefont
			{Valentine}},\ }\bibfield  {title} {\bibinfo {title} {Low-power
			electrochemical modulation of silicon-based metasurfaces},\ }\href
	{https://doi.org/10.1021/acsphotonics.3c01224} {\bibfield  {journal}
		{\bibinfo  {journal} {ACS Photonics}\ }\textbf {\bibinfo {volume} {11}},\
		\bibinfo {pages} {445} (\bibinfo {year} {2024})}\BibitemShut {NoStop}%
	\bibitem [{\citenamefont {Lu}\ \emph {et~al.}(2021)\citenamefont {Lu},
		\citenamefont {Dong}, \citenamefont {Tijiptoharsono}, \citenamefont {Ng},
		\citenamefont {Wang}, \citenamefont {Rezaei}, \citenamefont {Wang},
		\citenamefont {Leong}, \citenamefont {Lim}, \citenamefont {Yang},\ and\
		\citenamefont {Simpson}}]{Lu2021}%
	\BibitemOpen
	\bibfield  {author} {\bibinfo {author} {\bibfnamefont {L.}~\bibnamefont
			{Lu}}, \bibinfo {author} {\bibfnamefont {Z.}~\bibnamefont {Dong}}, \bibinfo
		{author} {\bibfnamefont {F.}~\bibnamefont {Tijiptoharsono}}, \bibinfo
		{author} {\bibfnamefont {R.~J.~H.}\ \bibnamefont {Ng}}, \bibinfo {author}
		{\bibfnamefont {H.}~\bibnamefont {Wang}}, \bibinfo {author} {\bibfnamefont
			{S.~D.}\ \bibnamefont {Rezaei}}, \bibinfo {author} {\bibfnamefont
			{Y.}~\bibnamefont {Wang}}, \bibinfo {author} {\bibfnamefont {H.~S.}\
			\bibnamefont {Leong}}, \bibinfo {author} {\bibfnamefont {P.~C.}\ \bibnamefont
			{Lim}}, \bibinfo {author} {\bibfnamefont {J.~K.~W.}\ \bibnamefont {Yang}},\
		and\ \bibinfo {author} {\bibfnamefont {R.~E.}\ \bibnamefont {Simpson}},\
	}\bibfield  {title} {\bibinfo {title} {Reversible tuning of Mie resonances in
			the visible spectrum},\ }\href {https://doi.org/10.1021/acsnano.1c07114}
	{\bibfield  {journal} {\bibinfo  {journal} {ACS Nano}\ }\textbf {\bibinfo
			{volume} {15}},\ \bibinfo {pages} {19722} (\bibinfo {year}
		{2021})}\BibitemShut {NoStop}%
	\bibitem [{\citenamefont {Quan}\ \emph {et~al.}(2025)\citenamefont {Quan},
		\citenamefont {Gu}, \citenamefont {Fu}, \citenamefont {Liu}, \citenamefont
		{Xu}, \citenamefont {Guo}, \citenamefont {Zhu},\ and\ \citenamefont
		{Zhang}}]{Quan2025}%
	\BibitemOpen
	\bibfield  {author} {\bibinfo {author} {\bibfnamefont {C.}~\bibnamefont
			{Quan}}, \bibinfo {author} {\bibfnamefont {S.}~\bibnamefont {Gu}}, \bibinfo
		{author} {\bibfnamefont {T.}~\bibnamefont {Fu}}, \bibinfo {author}
		{\bibfnamefont {P.}~\bibnamefont {Liu}}, \bibinfo {author} {\bibfnamefont
			{W.}~\bibnamefont {Xu}}, \bibinfo {author} {\bibfnamefont {C.}~\bibnamefont
			{Guo}}, \bibinfo {author} {\bibfnamefont {Z.}~\bibnamefont {Zhu}},\ and\
		\bibinfo {author} {\bibfnamefont {J.}~\bibnamefont {Zhang}},\ }\bibfield
	{title} {\bibinfo {title} {A non-volatile switchable infrared stealth
			metafilm with GST},\ }\href {https://doi.org/10.37188/lam.2025.016}
	{\bibfield  {journal} {\bibinfo  {journal} {Light Adv. Manuf.}\ }\textbf
		{\bibinfo {volume} {6}},\ \bibinfo {pages} {16} (\bibinfo {year}
		{2025})}\BibitemShut {NoStop}%
	\bibitem [{\citenamefont {Zou}\ \emph {et~al.}(2019)\citenamefont {Zou},
		\citenamefont {Komar}, \citenamefont {Fasold}, \citenamefont {Bohn},
		\citenamefont {Muravsky}, \citenamefont {Murauski}, \citenamefont {Pertsch},
		\citenamefont {Neshev},\ and\ \citenamefont {Staude}}]{Zou2019}%
	\BibitemOpen
	\bibfield  {author} {\bibinfo {author} {\bibfnamefont {C.}~\bibnamefont
			{Zou}}, \bibinfo {author} {\bibfnamefont {A.}~\bibnamefont {Komar}}, \bibinfo
		{author} {\bibfnamefont {S.}~\bibnamefont {Fasold}}, \bibinfo {author}
		{\bibfnamefont {J.}~\bibnamefont {Bohn}}, \bibinfo {author} {\bibfnamefont
			{A.~A.}\ \bibnamefont {Muravsky}}, \bibinfo {author} {\bibfnamefont {A.~A.}\
			\bibnamefont {Murauski}}, \bibinfo {author} {\bibfnamefont {T.}~\bibnamefont
			{Pertsch}}, \bibinfo {author} {\bibfnamefont {D.~N.}\ \bibnamefont
			{Neshev}},\ and\ \bibinfo {author} {\bibfnamefont {I.}~\bibnamefont
			{Staude}},\ }\bibfield  {title} {\bibinfo {title} {Electrically tunable
			transparent displays for visible light based on dielectric metasurfaces},\
	}\href {https://doi.org/10.1021/acsphotonics.9b00301} {\bibfield  {journal}
		{\bibinfo  {journal} {ACS Photonics}\ }\textbf {\bibinfo {volume} {6}},\
		\bibinfo {pages} {1533} (\bibinfo {year} {2019})}\BibitemShut {NoStop}%
	\bibitem [{\citenamefont {Beddoe}\ \emph {et~al.}(2025)\citenamefont {Beddoe},
		\citenamefont {Walden}, \citenamefont {Miljevic}, \citenamefont {Pidgayko},
		\citenamefont {Zou}, \citenamefont {Minovich}, \citenamefont {Barreda},
		\citenamefont {Pertsch},\ and\ \citenamefont {Staude}}]{Beddoe2025}%
	\BibitemOpen
	\bibfield  {author} {\bibinfo {author} {\bibfnamefont {M.}~\bibnamefont
			{Beddoe}}, \bibinfo {author} {\bibfnamefont {S.~L.}\ \bibnamefont {Walden}},
		\bibinfo {author} {\bibfnamefont {S.}~\bibnamefont {Miljevic}}, \bibinfo
		{author} {\bibfnamefont {D.}~\bibnamefont {Pidgayko}}, \bibinfo {author}
		{\bibfnamefont {C.}~\bibnamefont {Zou}}, \bibinfo {author} {\bibfnamefont
			{A.~E.}\ \bibnamefont {Minovich}}, \bibinfo {author} {\bibfnamefont
			{A.}~\bibnamefont {Barreda}}, \bibinfo {author} {\bibfnamefont
			{T.}~\bibnamefont {Pertsch}},\ and\ \bibinfo {author} {\bibfnamefont
			{I.}~\bibnamefont {Staude}},\ }\bibfield  {title} {\bibinfo {title}
		{Spatially controlled all-optical switching of liquid-crystal-empowered
			metasurfaces},\ }\href {https://doi.org/10.1021/acsphotonics.4c02029}
	{\bibfield  {journal} {\bibinfo  {journal} {ACS Photonics}\ }\textbf
		{\bibinfo {volume} {12}},\ \bibinfo {pages} {963} (\bibinfo {year}
		{2025})}\BibitemShut {NoStop}%
	\bibitem [{\citenamefont {Zhang}\ \emph {et~al.}(2023)\citenamefont {Zhang},
		\citenamefont {Sun}, \citenamefont {Yu}, \citenamefont {Deng}, \citenamefont
		{Qiu}, \citenamefont {Wang},\ and\ \citenamefont {Qiu}}]{Zhang2023a}%
	\BibitemOpen
	\bibfield  {author} {\bibinfo {author} {\bibfnamefont {L.}~\bibnamefont
			{Zhang}}, \bibinfo {author} {\bibfnamefont {X.}~\bibnamefont {Sun}}, \bibinfo
		{author} {\bibfnamefont {H.}~\bibnamefont {Yu}}, \bibinfo {author}
		{\bibfnamefont {N.}~\bibnamefont {Deng}}, \bibinfo {author} {\bibfnamefont
			{F.}~\bibnamefont {Qiu}}, \bibinfo {author} {\bibfnamefont {J.}~\bibnamefont
			{Wang}},\ and\ \bibinfo {author} {\bibfnamefont {M.}~\bibnamefont {Qiu}},\
	}\bibfield  {title} {\bibinfo {title} {Plasmonic metafibers electro-optic
			modulators},\ }\href {https://doi.org/10.1038/s41377-023-01255-7} {\bibfield
		{journal} {\bibinfo  {journal} {Light Sci. Appl.}\ }\textbf {\bibinfo
			{volume} {12}},\ \bibinfo {pages} {198} (\bibinfo {year} {2023})}\BibitemShut
	{NoStop}%
	\bibitem [{\citenamefont {Benea-Chelmus}\ \emph {et~al.}(2022)\citenamefont
		{Benea-Chelmus}, \citenamefont {Mason}, \citenamefont {Meretska},
		\citenamefont {Elder}, \citenamefont {Kazakov}, \citenamefont {Shams-Ansari},
		\citenamefont {Dalton},\ and\ \citenamefont {Capasso}}]{BeneaChelmus2022}%
	\BibitemOpen
	\bibfield  {author} {\bibinfo {author} {\bibfnamefont {I.-C.}\ \bibnamefont
			{Benea-Chelmus}}, \bibinfo {author} {\bibfnamefont {S.}~\bibnamefont
			{Mason}}, \bibinfo {author} {\bibfnamefont {M.~L.}\ \bibnamefont {Meretska}},
		\bibinfo {author} {\bibfnamefont {D.~L.}\ \bibnamefont {Elder}}, \bibinfo
		{author} {\bibfnamefont {D.}~\bibnamefont {Kazakov}}, \bibinfo {author}
		{\bibfnamefont {A.}~\bibnamefont {Shams-Ansari}}, \bibinfo {author}
		{\bibfnamefont {L.~R.}\ \bibnamefont {Dalton}},\ and\ \bibinfo {author}
		{\bibfnamefont {F.}~\bibnamefont {Capasso}},\ }\bibfield  {title} {\bibinfo
		{title} {Gigahertz free-space electro-optic modulators based on Mie
			resonances},\ }\href {https://doi.org/10.1038/s41467-022-30451-z} {\bibfield
		{journal} {\bibinfo  {journal} {Nat. Commun.}\ }\textbf {\bibinfo {volume}
			{13}},\ \bibinfo {pages} {3170} (\bibinfo {year} {2022})}\BibitemShut
	{NoStop}%
	\bibitem [{\citenamefont {Zheng}\ \emph {et~al.}(2024)\citenamefont {Zheng},
		\citenamefont {Gu}, \citenamefont {Kwon}, \citenamefont {Roberts},\ and\
		\citenamefont {Faraon}}]{Zheng2024}%
	\BibitemOpen
	\bibfield  {author} {\bibinfo {author} {\bibfnamefont {T.}~\bibnamefont
			{Zheng}}, \bibinfo {author} {\bibfnamefont {Y.}~\bibnamefont {Gu}}, \bibinfo
		{author} {\bibfnamefont {H.}~\bibnamefont {Kwon}}, \bibinfo {author}
		{\bibfnamefont {G.}~\bibnamefont {Roberts}},\ and\ \bibinfo {author}
		{\bibfnamefont {A.}~\bibnamefont {Faraon}},\ }\bibfield  {title} {\bibinfo
		{title} {Dynamic light manipulation via silicon-organic slot metasurfaces},\
	}\href {https://doi.org/10.1038/s41467-024-45544-0} {\bibfield  {journal}
		{\bibinfo  {journal} {Nat. Commun.}\ }\textbf {\bibinfo {volume} {15}},\
		\bibinfo {pages} {1557} (\bibinfo {year} {2024})}\BibitemShut {NoStop}%
	\bibitem [{\citenamefont {Chen}\ \emph {et~al.}(2022)\citenamefont {Chen},
		\citenamefont {Li}, \citenamefont {Ng}, \citenamefont {Lin}, \citenamefont
		{Zhou}, \citenamefont {Fu}, \citenamefont {Lee}, \citenamefont {Yu},
		\citenamefont {Liu},\ and\ \citenamefont {Danner}}]{Chen2022}%
	\BibitemOpen
	\bibfield  {author} {\bibinfo {author} {\bibfnamefont {G.}~\bibnamefont
			{Chen}}, \bibinfo {author} {\bibfnamefont {N.}~\bibnamefont {Li}}, \bibinfo
		{author} {\bibfnamefont {J.~D.}\ \bibnamefont {Ng}}, \bibinfo {author}
		{\bibfnamefont {H.-L.}\ \bibnamefont {Lin}}, \bibinfo {author} {\bibfnamefont
			{Y.}~\bibnamefont {Zhou}}, \bibinfo {author} {\bibfnamefont {Y.~H.}\
			\bibnamefont {Fu}}, \bibinfo {author} {\bibfnamefont {L.~Y.~T.}\ \bibnamefont
			{Lee}}, \bibinfo {author} {\bibfnamefont {Y.}~\bibnamefont {Yu}}, \bibinfo
		{author} {\bibfnamefont {A.-Q.}\ \bibnamefont {Liu}},\ and\ \bibinfo {author}
		{\bibfnamefont {A.~J.}\ \bibnamefont {Danner}},\ }\bibfield  {title}
	{\bibinfo {title} {Advances in lithium niobate photonics: development status
			and perspectives},\ }\href {https://doi.org/10.1117/1.ap.4.3.034003}
	{\bibfield  {journal} {\bibinfo  {journal} {Adv. Photonics}\ }\textbf
		{\bibinfo {volume} {4}},\ \bibinfo {pages} {034003} (\bibinfo {year}
		{2022})}\BibitemShut {NoStop}%
	\bibitem [{\citenamefont {Fedotova}\ \emph {et~al.}(2022)\citenamefont
		{Fedotova}, \citenamefont {Carletti}, \citenamefont {Zilli}, \citenamefont
		{Setzpfandt}, \citenamefont {Staude}, \citenamefont {Toma}, \citenamefont
		{Finazzi}, \citenamefont {De~Angelis}, \citenamefont {Pertsch}, \citenamefont
		{Neshev} \emph {et~al.}}]{Fedotova2022}%
	\BibitemOpen
	\bibfield  {author} {\bibinfo {author} {\bibfnamefont {A.}~\bibnamefont
			{Fedotova}}, \bibinfo {author} {\bibfnamefont {L.}~\bibnamefont {Carletti}},
		\bibinfo {author} {\bibfnamefont {A.}~\bibnamefont {Zilli}}, \bibinfo
		{author} {\bibfnamefont {F.}~\bibnamefont {Setzpfandt}}, \bibinfo {author}
		{\bibfnamefont {I.}~\bibnamefont {Staude}}, \bibinfo {author} {\bibfnamefont
			{A.}~\bibnamefont {Toma}}, \bibinfo {author} {\bibfnamefont {M.}~\bibnamefont
			{Finazzi}}, \bibinfo {author} {\bibfnamefont {C.}~\bibnamefont {De~Angelis}},
		\bibinfo {author} {\bibfnamefont {T.}~\bibnamefont {Pertsch}}, \bibinfo
		{author} {\bibfnamefont {D.~N.}\ \bibnamefont {Neshev}}, \emph {et~al.},\
	}\bibfield  {title} {\bibinfo {title} {Lithium niobate meta-optics},\ }\href
	{https://doi.org/10.1021/acsphotonics.2c00835} {\bibfield  {journal}
		{\bibinfo  {journal} {ACS Photonics}\ }\textbf {\bibinfo {volume} {9}},\
		\bibinfo {pages} {3745} (\bibinfo {year} {2022})}\BibitemShut {NoStop}%
	\bibitem [{\citenamefont {Boes}\ \emph {et~al.}(2023)\citenamefont {Boes},
		\citenamefont {Chang}, \citenamefont {Langrock}, \citenamefont {Yu},
		\citenamefont {Zhang}, \citenamefont {Lin}, \citenamefont {Lončar},
		\citenamefont {Fejer}, \citenamefont {Bowers},\ and\ \citenamefont
		{Mitchell}}]{Boes2023}%
	\BibitemOpen
	\bibfield  {author} {\bibinfo {author} {\bibfnamefont {A.}~\bibnamefont
			{Boes}}, \bibinfo {author} {\bibfnamefont {L.}~\bibnamefont {Chang}},
		\bibinfo {author} {\bibfnamefont {C.}~\bibnamefont {Langrock}}, \bibinfo
		{author} {\bibfnamefont {M.}~\bibnamefont {Yu}}, \bibinfo {author}
		{\bibfnamefont {M.}~\bibnamefont {Zhang}}, \bibinfo {author} {\bibfnamefont
			{Q.}~\bibnamefont {Lin}}, \bibinfo {author} {\bibfnamefont {M.}~\bibnamefont
			{Lončar}}, \bibinfo {author} {\bibfnamefont {M.}~\bibnamefont {Fejer}},
		\bibinfo {author} {\bibfnamefont {J.}~\bibnamefont {Bowers}},\ and\ \bibinfo
		{author} {\bibfnamefont {A.}~\bibnamefont {Mitchell}},\ }\bibfield  {title}
	{\bibinfo {title} {Lithium niobate photonics: Unlocking the electromagnetic
			spectrum},\ }\href {https://doi.org/10.1126/science.abj4396} {\bibfield
		{journal} {\bibinfo  {journal} {Science}\ }\textbf {\bibinfo {volume}
			{379}},\ \bibinfo {pages} {eabj4396} (\bibinfo {year} {2023})}\BibitemShut
	{NoStop}%
	\bibitem [{\citenamefont {Fedotova}\ \emph {et~al.}(2020)\citenamefont
		{Fedotova}, \citenamefont {Younesi}, \citenamefont {Sautter}, \citenamefont
		{Vaskin}, \citenamefont {Löchner}, \citenamefont {Steinert}, \citenamefont
		{Geiss}, \citenamefont {Pertsch}, \citenamefont {Staude},\ and\ \citenamefont
		{Setzpfandt}}]{Fedotova2020}%
	\BibitemOpen
	\bibfield  {author} {\bibinfo {author} {\bibfnamefont {A.}~\bibnamefont
			{Fedotova}}, \bibinfo {author} {\bibfnamefont {M.}~\bibnamefont {Younesi}},
		\bibinfo {author} {\bibfnamefont {J.}~\bibnamefont {Sautter}}, \bibinfo
		{author} {\bibfnamefont {A.}~\bibnamefont {Vaskin}}, \bibinfo {author}
		{\bibfnamefont {F.~J.}\ \bibnamefont {Löchner}}, \bibinfo {author}
		{\bibfnamefont {M.}~\bibnamefont {Steinert}}, \bibinfo {author}
		{\bibfnamefont {R.}~\bibnamefont {Geiss}}, \bibinfo {author} {\bibfnamefont
			{T.}~\bibnamefont {Pertsch}}, \bibinfo {author} {\bibfnamefont
			{I.}~\bibnamefont {Staude}},\ and\ \bibinfo {author} {\bibfnamefont
			{F.}~\bibnamefont {Setzpfandt}},\ }\bibfield  {title} {\bibinfo {title}
		{Second-harmonic generation in resonant nonlinear metasurfaces based on
			lithium niobate},\ }\href {https://doi.org/10.1021/acs.nanolett.0c03290}
	{\bibfield  {journal} {\bibinfo  {journal} {Nano Lett.}\ }\textbf {\bibinfo
			{volume} {20}},\ \bibinfo {pages} {8608} (\bibinfo {year}
		{2020})}\BibitemShut {NoStop}%
	\bibitem [{\citenamefont {Yuan}\ \emph {et~al.}(2021)\citenamefont {Yuan},
		\citenamefont {Wu}, \citenamefont {Dang}, \citenamefont {Zeng}, \citenamefont
		{Qi}, \citenamefont {Guo}, \citenamefont {Ren},\ and\ \citenamefont
		{Xia}}]{Yuan2021}%
	\BibitemOpen
	\bibfield  {author} {\bibinfo {author} {\bibfnamefont {S.}~\bibnamefont
			{Yuan}}, \bibinfo {author} {\bibfnamefont {Y.}~\bibnamefont {Wu}}, \bibinfo
		{author} {\bibfnamefont {Z.}~\bibnamefont {Dang}}, \bibinfo {author}
		{\bibfnamefont {C.}~\bibnamefont {Zeng}}, \bibinfo {author} {\bibfnamefont
			{X.}~\bibnamefont {Qi}}, \bibinfo {author} {\bibfnamefont {G.}~\bibnamefont
			{Guo}}, \bibinfo {author} {\bibfnamefont {X.}~\bibnamefont {Ren}},\ and\
		\bibinfo {author} {\bibfnamefont {J.}~\bibnamefont {Xia}},\ }\bibfield
	{title} {\bibinfo {title} {Strongly enhanced second harmonic generation in a
			thin film lithium niobate heterostructure cavity},\ }\href
	{https://doi.org/10.1103/physrevlett.127.153901} {\bibfield  {journal}
		{\bibinfo  {journal} {Phys. Rev. Lett.}\ }\textbf {\bibinfo {volume} {127}},\
		\bibinfo {pages} {153901} (\bibinfo {year} {2021})}\BibitemShut {NoStop}%
	\bibitem [{\citenamefont {Ma}\ \emph {et~al.}(2021)\citenamefont {Ma},
		\citenamefont {Xie}, \citenamefont {Chen}, \citenamefont {Chen},
		\citenamefont {Wu}, \citenamefont {Liu}, \citenamefont {Chen}, \citenamefont
		{Cai}, \citenamefont {Ren},\ and\ \citenamefont {Xu}}]{Ma2021a}%
	\BibitemOpen
	\bibfield  {author} {\bibinfo {author} {\bibfnamefont {J.}~\bibnamefont
			{Ma}}, \bibinfo {author} {\bibfnamefont {F.}~\bibnamefont {Xie}}, \bibinfo
		{author} {\bibfnamefont {W.}~\bibnamefont {Chen}}, \bibinfo {author}
		{\bibfnamefont {J.}~\bibnamefont {Chen}}, \bibinfo {author} {\bibfnamefont
			{W.}~\bibnamefont {Wu}}, \bibinfo {author} {\bibfnamefont {W.}~\bibnamefont
			{Liu}}, \bibinfo {author} {\bibfnamefont {Y.}~\bibnamefont {Chen}}, \bibinfo
		{author} {\bibfnamefont {W.}~\bibnamefont {Cai}}, \bibinfo {author}
		{\bibfnamefont {M.}~\bibnamefont {Ren}},\ and\ \bibinfo {author}
		{\bibfnamefont {J.}~\bibnamefont {Xu}},\ }\bibfield  {title} {\bibinfo
		{title} {Nonlinear lithium niobate metasurfaces for second harmonic
			generation},\ }\href {https://doi.org/10.1002/lpor.202000521} {\bibfield
		{journal} {\bibinfo  {journal} {Laser Photonics Rev.}\ }\textbf {\bibinfo
			{volume} {15}},\ \bibinfo {pages} {2000521} (\bibinfo {year}
		{2021})}\BibitemShut {NoStop}%
	\bibitem [{\citenamefont {Huang}\ \emph {et~al.}(2022)\citenamefont {Huang},
		\citenamefont {Luo}, \citenamefont {Feng}, \citenamefont {Zhang},
		\citenamefont {Li}, \citenamefont {Qiu}, \citenamefont {Guan}, \citenamefont
		{Xu}, \citenamefont {Li},\ and\ \citenamefont {Lu}}]{Huang2022}%
	\BibitemOpen
	\bibfield  {author} {\bibinfo {author} {\bibfnamefont {Z.}~\bibnamefont
			{Huang}}, \bibinfo {author} {\bibfnamefont {K.}~\bibnamefont {Luo}}, \bibinfo
		{author} {\bibfnamefont {Z.}~\bibnamefont {Feng}}, \bibinfo {author}
		{\bibfnamefont {Z.}~\bibnamefont {Zhang}}, \bibinfo {author} {\bibfnamefont
			{Y.}~\bibnamefont {Li}}, \bibinfo {author} {\bibfnamefont {W.}~\bibnamefont
			{Qiu}}, \bibinfo {author} {\bibfnamefont {H.}~\bibnamefont {Guan}}, \bibinfo
		{author} {\bibfnamefont {Y.}~\bibnamefont {Xu}}, \bibinfo {author}
		{\bibfnamefont {X.}~\bibnamefont {Li}},\ and\ \bibinfo {author}
		{\bibfnamefont {H.}~\bibnamefont {Lu}},\ }\bibfield  {title} {\bibinfo
		{title} {Resonant enhancement of second harmonic generation in etchless thin
			film lithium niobate heteronanostructure},\ }\href
	{https://doi.org/10.1007/s11433-022-1937-8} {\bibfield  {journal} {\bibinfo
			{journal} {Sci. China Phys. Mech. Astron.}\ }\textbf {\bibinfo {volume}
			{65}},\ \bibinfo {pages} {104211} (\bibinfo {year} {2022})}\BibitemShut
	{NoStop}%
	\bibitem [{\citenamefont {Qu}\ \emph {et~al.}(2022)\citenamefont {Qu},
		\citenamefont {Bai}, \citenamefont {Jin}, \citenamefont {Liu}, \citenamefont
		{Wu}, \citenamefont {Gao}, \citenamefont {Li}, \citenamefont {Cai},
		\citenamefont {Ren},\ and\ \citenamefont {Xu}}]{Qu2022}%
	\BibitemOpen
	\bibfield  {author} {\bibinfo {author} {\bibfnamefont {L.}~\bibnamefont
			{Qu}}, \bibinfo {author} {\bibfnamefont {L.}~\bibnamefont {Bai}}, \bibinfo
		{author} {\bibfnamefont {C.}~\bibnamefont {Jin}}, \bibinfo {author}
		{\bibfnamefont {Q.}~\bibnamefont {Liu}}, \bibinfo {author} {\bibfnamefont
			{W.}~\bibnamefont {Wu}}, \bibinfo {author} {\bibfnamefont {B.}~\bibnamefont
			{Gao}}, \bibinfo {author} {\bibfnamefont {J.}~\bibnamefont {Li}}, \bibinfo
		{author} {\bibfnamefont {W.}~\bibnamefont {Cai}}, \bibinfo {author}
		{\bibfnamefont {M.}~\bibnamefont {Ren}},\ and\ \bibinfo {author}
		{\bibfnamefont {J.}~\bibnamefont {Xu}},\ }\bibfield  {title} {\bibinfo
		{title} {Giant second harmonic generation from membrane metasurfaces},\
	}\href {https://doi.org/10.1021/acs.nanolett.2c03811} {\bibfield  {journal}
		{\bibinfo  {journal} {Nano Lett.}\ }\textbf {\bibinfo {volume} {22}},\
		\bibinfo {pages} {9652} (\bibinfo {year} {2022})}\BibitemShut {NoStop}%
	\bibitem [{\citenamefont {Qu}\ \emph {et~al.}(2023)\citenamefont {Qu},
		\citenamefont {Gu}, \citenamefont {Li}, \citenamefont {Qin}, \citenamefont
		{Zhang}, \citenamefont {Zhang}, \citenamefont {Zhao}, \citenamefont {Liu},
		\citenamefont {Jin}, \citenamefont {Wang}, \citenamefont {Wu}, \citenamefont
		{Cai}, \citenamefont {Liu}, \citenamefont {Ren},\ and\ \citenamefont
		{Xu}}]{Qu2023}%
	\BibitemOpen
	\bibfield  {author} {\bibinfo {author} {\bibfnamefont {L.}~\bibnamefont
			{Qu}}, \bibinfo {author} {\bibfnamefont {Z.}~\bibnamefont {Gu}}, \bibinfo
		{author} {\bibfnamefont {C.}~\bibnamefont {Li}}, \bibinfo {author}
		{\bibfnamefont {Y.}~\bibnamefont {Qin}}, \bibinfo {author} {\bibfnamefont
			{Y.}~\bibnamefont {Zhang}}, \bibinfo {author} {\bibfnamefont
			{D.}~\bibnamefont {Zhang}}, \bibinfo {author} {\bibfnamefont
			{J.}~\bibnamefont {Zhao}}, \bibinfo {author} {\bibfnamefont {Q.}~\bibnamefont
			{Liu}}, \bibinfo {author} {\bibfnamefont {C.}~\bibnamefont {Jin}}, \bibinfo
		{author} {\bibfnamefont {L.}~\bibnamefont {Wang}}, \bibinfo {author}
		{\bibfnamefont {W.}~\bibnamefont {Wu}}, \bibinfo {author} {\bibfnamefont
			{W.}~\bibnamefont {Cai}}, \bibinfo {author} {\bibfnamefont {H.}~\bibnamefont
			{Liu}}, \bibinfo {author} {\bibfnamefont {M.}~\bibnamefont {Ren}},\ and\
		\bibinfo {author} {\bibfnamefont {J.}~\bibnamefont {Xu}},\ }\bibfield
	{title} {\bibinfo {title} {Bright second harmonic emission from photonic
			crystal vertical cavity},\ }\href {https://doi.org/10.1002/adfm.202308484}
	{\bibfield  {journal} {\bibinfo  {journal} {Adv. Funct. Mater.}\ }\textbf
		{\bibinfo {volume} {33}},\ \bibinfo {pages} {2308484} (\bibinfo {year}
		{2023})}\BibitemShut {NoStop}%
	\bibitem [{\citenamefont {Jiang}\ \emph {et~al.}(2024)\citenamefont {Jiang},
		\citenamefont {Sun}, \citenamefont {Jia}, \citenamefont {Cai}, \citenamefont
		{Levy},\ and\ \citenamefont {Han}}]{Jiang2024}%
	\BibitemOpen
	\bibfield  {author} {\bibinfo {author} {\bibfnamefont {H.}~\bibnamefont
			{Jiang}}, \bibinfo {author} {\bibfnamefont {K.}~\bibnamefont {Sun}}, \bibinfo
		{author} {\bibfnamefont {Y.}~\bibnamefont {Jia}}, \bibinfo {author}
		{\bibfnamefont {Y.}~\bibnamefont {Cai}}, \bibinfo {author} {\bibfnamefont
			{U.}~\bibnamefont {Levy}},\ and\ \bibinfo {author} {\bibfnamefont
			{Z.}~\bibnamefont {Han}},\ }\bibfield  {title} {\bibinfo {title} {Tunable
			second harmonic generation with large enhancement in a nonlocal
			all‐dielectric metasurface over a broad spectral range},\ }\href
	{https://doi.org/10.1002/adom.202303229} {\bibfield  {journal} {\bibinfo
			{journal} {Adv. Opt. Mater.}\ }\textbf {\bibinfo {volume} {12}},\ \bibinfo
		{pages} {2303229} (\bibinfo {year} {2024})}\BibitemShut {NoStop}%
	\bibitem [{\citenamefont {Qu}\ \emph {et~al.}(2025)\citenamefont {Qu},
		\citenamefont {Wu}, \citenamefont {Cai}, \citenamefont {Ren},\ and\
		\citenamefont {Xu}}]{Qu2025}%
	\BibitemOpen
	\bibfield  {author} {\bibinfo {author} {\bibfnamefont {L.}~\bibnamefont
			{Qu}}, \bibinfo {author} {\bibfnamefont {W.}~\bibnamefont {Wu}}, \bibinfo
		{author} {\bibfnamefont {W.}~\bibnamefont {Cai}}, \bibinfo {author}
		{\bibfnamefont {M.}~\bibnamefont {Ren}},\ and\ \bibinfo {author}
		{\bibfnamefont {J.}~\bibnamefont {Xu}},\ }\bibfield  {title} {\bibinfo
		{title} {Second harmonic generation in lithium niobate on insulator},\ }\href
	{https://doi.org/10.1002/lpor.202401928} {\bibfield  {journal} {\bibinfo
			{journal} {Laser Photonics Rev.}\ ,\ \bibinfo {pages} {2401928}} (\bibinfo
		{year} {2025})}\BibitemShut {NoStop}%
	\bibitem [{\citenamefont {He}\ \emph {et~al.}(2024)\citenamefont {He},
		\citenamefont {Qu}, \citenamefont {Wu}, \citenamefont {Liu}, \citenamefont
		{Jin}, \citenamefont {Wang}, \citenamefont {You}, \citenamefont {Liu},
		\citenamefont {Bai}, \citenamefont {Gu} \emph {et~al.}}]{He2024}%
	\BibitemOpen
	\bibfield  {author} {\bibinfo {author} {\bibfnamefont {Z.}~\bibnamefont
			{He}}, \bibinfo {author} {\bibfnamefont {L.}~\bibnamefont {Qu}}, \bibinfo
		{author} {\bibfnamefont {W.}~\bibnamefont {Wu}}, \bibinfo {author}
		{\bibfnamefont {J.}~\bibnamefont {Liu}}, \bibinfo {author} {\bibfnamefont
			{C.}~\bibnamefont {Jin}}, \bibinfo {author} {\bibfnamefont {C.}~\bibnamefont
			{Wang}}, \bibinfo {author} {\bibfnamefont {J.}~\bibnamefont {You}}, \bibinfo
		{author} {\bibfnamefont {W.}~\bibnamefont {Liu}}, \bibinfo {author}
		{\bibfnamefont {L.}~\bibnamefont {Bai}}, \bibinfo {author} {\bibfnamefont
			{Z.}~\bibnamefont {Gu}}, \emph {et~al.},\ }\bibfield  {title} {\bibinfo
		{title} {Electro-optically modulated nonlinear metasurfaces},\ }\href
	{https://doi.org/10.1021/acs.nanolett.4c03369} {\bibfield  {journal}
		{\bibinfo  {journal} {Nano Lett.}\ }\textbf {\bibinfo {volume} {24}},\
		\bibinfo {pages} {14215} (\bibinfo {year} {2024})}\BibitemShut {NoStop}%
	\bibitem [{\citenamefont {Di~Francescantonio}\ \emph
		{et~al.}(2024)\citenamefont {Di~Francescantonio}, \citenamefont {Sabatti},
		\citenamefont {Weigand}, \citenamefont {Bailly}, \citenamefont {Vincenti},
		\citenamefont {Carletti}, \citenamefont {Kellner}, \citenamefont {Zilli},
		\citenamefont {Finazzi}, \citenamefont {Celebrano} \emph
		{et~al.}}]{DiFrancescantonio2024}%
	\BibitemOpen
	\bibfield  {author} {\bibinfo {author} {\bibfnamefont {A.}~\bibnamefont
			{Di~Francescantonio}}, \bibinfo {author} {\bibfnamefont {A.}~\bibnamefont
			{Sabatti}}, \bibinfo {author} {\bibfnamefont {H.}~\bibnamefont {Weigand}},
		\bibinfo {author} {\bibfnamefont {E.}~\bibnamefont {Bailly}}, \bibinfo
		{author} {\bibfnamefont {M.~A.}\ \bibnamefont {Vincenti}}, \bibinfo {author}
		{\bibfnamefont {L.}~\bibnamefont {Carletti}}, \bibinfo {author}
		{\bibfnamefont {J.}~\bibnamefont {Kellner}}, \bibinfo {author} {\bibfnamefont
			{A.}~\bibnamefont {Zilli}}, \bibinfo {author} {\bibfnamefont
			{M.}~\bibnamefont {Finazzi}}, \bibinfo {author} {\bibfnamefont
			{M.}~\bibnamefont {Celebrano}}, \emph {et~al.},\ }\bibfield  {title}
	{\bibinfo {title} {Efficient GHz electro-optical modulation with a nonlocal
			lithium niobate metasurface in the linear and nonlinear regime},\ }\href
	{https://doi.org/10.48550/arXiv.2412.03422} {\bibfield  {journal} {\bibinfo
			{journal} {arXiv preprint arXiv:2412.03422}\ } (\bibinfo {year}
		{2024})}\BibitemShut {NoStop}%
	\bibitem [{\citenamefont {Kanyang}\ \emph {et~al.}(2025)\citenamefont
		{Kanyang}, \citenamefont {Fang}, \citenamefont {Wang}, \citenamefont {Liu},
		\citenamefont {Hao},\ and\ \citenamefont {Han}}]{Kanyang2025}%
	\BibitemOpen
	\bibfield  {author} {\bibinfo {author} {\bibfnamefont {R.}~\bibnamefont
			{Kanyang}}, \bibinfo {author} {\bibfnamefont {C.}~\bibnamefont {Fang}},
		\bibinfo {author} {\bibfnamefont {Y.}~\bibnamefont {Wang}}, \bibinfo {author}
		{\bibfnamefont {Y.}~\bibnamefont {Liu}}, \bibinfo {author} {\bibfnamefont
			{Y.}~\bibnamefont {Hao}},\ and\ \bibinfo {author} {\bibfnamefont
			{G.}~\bibnamefont {Han}},\ }\bibfield  {title} {\bibinfo {title} {Efficient
			tunable nonlinearity in high $Q$-factor lithium niobate metasurface harnessing
			electric-field-induced second harmonic generation},\ }\href
	{https://doi.org/10.1364/OE.554716} {\bibfield  {journal} {\bibinfo
			{journal} {Opt. Express}\ }\textbf {\bibinfo {volume} {33}},\ \bibinfo
		{pages} {7247} (\bibinfo {year} {2025})}\BibitemShut {NoStop}%
	\bibitem [{\citenamefont {Hsu}\ \emph {et~al.}(2016)\citenamefont {Hsu},
		\citenamefont {Zhen}, \citenamefont {Stone}, \citenamefont {Joannopoulos},\
		and\ \citenamefont {Soljačić}}]{Hsu2016}%
	\BibitemOpen
	\bibfield  {author} {\bibinfo {author} {\bibfnamefont {C.~W.}\ \bibnamefont
			{Hsu}}, \bibinfo {author} {\bibfnamefont {B.}~\bibnamefont {Zhen}}, \bibinfo
		{author} {\bibfnamefont {A.~D.}\ \bibnamefont {Stone}}, \bibinfo {author}
		{\bibfnamefont {J.~D.}\ \bibnamefont {Joannopoulos}},\ and\ \bibinfo {author}
		{\bibfnamefont {M.}~\bibnamefont {Soljačić}},\ }\bibfield  {title}
	{\bibinfo {title} {Bound states in the continuum},\ }\href
	{https://doi.org/10.1038/natrevmats.2016.48} {\bibfield  {journal} {\bibinfo
			{journal} {Nat. Rev. Mater.}\ }\textbf {\bibinfo {volume} {1}},\ \bibinfo
		{pages} {16048} (\bibinfo {year} {2016})}\BibitemShut {NoStop}%
	\bibitem [{\citenamefont {Hwang}\ \emph {et~al.}(2022)\citenamefont {Hwang},
		\citenamefont {Jeong}, \citenamefont {So}, \citenamefont {Kim},\ and\
		\citenamefont {Park}}]{Hwang2022}%
	\BibitemOpen
	\bibfield  {author} {\bibinfo {author} {\bibfnamefont {M.-S.}\ \bibnamefont
			{Hwang}}, \bibinfo {author} {\bibfnamefont {K.-Y.}\ \bibnamefont {Jeong}},
		\bibinfo {author} {\bibfnamefont {J.-P.}\ \bibnamefont {So}}, \bibinfo
		{author} {\bibfnamefont {K.-H.}\ \bibnamefont {Kim}},\ and\ \bibinfo {author}
		{\bibfnamefont {H.-G.}\ \bibnamefont {Park}},\ }\bibfield  {title} {\bibinfo
		{title} {Nanophotonic nonlinear and laser devices exploiting bound states in
			the continuum},\ }\href {https://doi.org/10.1038/s42005-022-00884-5}
	{\bibfield  {journal} {\bibinfo  {journal} {Commun. Phys.}\ }\textbf
		{\bibinfo {volume} {5}},\ \bibinfo {pages} {106} (\bibinfo {year}
		{2022})}\BibitemShut {NoStop}%
	\bibitem [{\citenamefont {Huang}\ \emph {et~al.}(2023)\citenamefont {Huang},
		\citenamefont {Xu}, \citenamefont {Powell}, \citenamefont {Padilla},\ and\
		\citenamefont {Miroshnichenko}}]{Huang2023}%
	\BibitemOpen
	\bibfield  {author} {\bibinfo {author} {\bibfnamefont {L.}~\bibnamefont
			{Huang}}, \bibinfo {author} {\bibfnamefont {L.}~\bibnamefont {Xu}}, \bibinfo
		{author} {\bibfnamefont {D.~A.}\ \bibnamefont {Powell}}, \bibinfo {author}
		{\bibfnamefont {W.~J.}\ \bibnamefont {Padilla}},\ and\ \bibinfo {author}
		{\bibfnamefont {A.~E.}\ \bibnamefont {Miroshnichenko}},\ }\bibfield  {title}
	{\bibinfo {title} {Resonant leaky modes in all-dielectric metasystems:
			Fundamentals and applications},\ }\href
	{https://doi.org/10.1016/j.physrep.2023.01.001} {\bibfield  {journal}
		{\bibinfo  {journal} {Phys. Rep.}\ }\textbf {\bibinfo {volume} {1008}},\
		\bibinfo {pages} {1} (\bibinfo {year} {2023})}\BibitemShut {NoStop}%
	\bibitem [{\citenamefont {Liu}\ \emph {et~al.}(2019)\citenamefont {Liu},
		\citenamefont {Xu}, \citenamefont {Lin}, \citenamefont {Xiang}, \citenamefont
		{Feng}, \citenamefont {Cao}, \citenamefont {Li}, \citenamefont {Lan},\ and\
		\citenamefont {Liu}}]{Liu2019}%
	\BibitemOpen
	\bibfield  {author} {\bibinfo {author} {\bibfnamefont {Z.}~\bibnamefont
			{Liu}}, \bibinfo {author} {\bibfnamefont {Y.}~\bibnamefont {Xu}}, \bibinfo
		{author} {\bibfnamefont {Y.}~\bibnamefont {Lin}}, \bibinfo {author}
		{\bibfnamefont {J.}~\bibnamefont {Xiang}}, \bibinfo {author} {\bibfnamefont
			{T.}~\bibnamefont {Feng}}, \bibinfo {author} {\bibfnamefont {Q.}~\bibnamefont
			{Cao}}, \bibinfo {author} {\bibfnamefont {J.}~\bibnamefont {Li}}, \bibinfo
		{author} {\bibfnamefont {S.}~\bibnamefont {Lan}},\ and\ \bibinfo {author}
		{\bibfnamefont {J.}~\bibnamefont {Liu}},\ }\bibfield  {title} {\bibinfo
		{title} {High-$Q$ quasibound states in the continuum for nonlinear
			metasurfaces},\ }\href {https://doi.org/10.1103/physrevlett.123.253901}
	{\bibfield  {journal} {\bibinfo  {journal} {Phys. Rev. Lett.}\ }\textbf
		{\bibinfo {volume} {123}},\ \bibinfo {pages} {253901} (\bibinfo {year}
		{2019})}\BibitemShut {NoStop}%
	\bibitem [{\citenamefont {Xu}\ \emph {et~al.}(2019)\citenamefont {Xu},
		\citenamefont {Zangeneh~Kamali}, \citenamefont {Huang}, \citenamefont
		{Rahmani}, \citenamefont {Smirnov}, \citenamefont {Camacho-Morales},
		\citenamefont {Ma}, \citenamefont {Zhang}, \citenamefont {Woolley},
		\citenamefont {Neshev} \emph {et~al.}}]{Xu2019}%
	\BibitemOpen
	\bibfield  {author} {\bibinfo {author} {\bibfnamefont {L.}~\bibnamefont
			{Xu}}, \bibinfo {author} {\bibfnamefont {K.}~\bibnamefont {Zangeneh~Kamali}},
		\bibinfo {author} {\bibfnamefont {L.}~\bibnamefont {Huang}}, \bibinfo
		{author} {\bibfnamefont {M.}~\bibnamefont {Rahmani}}, \bibinfo {author}
		{\bibfnamefont {A.}~\bibnamefont {Smirnov}}, \bibinfo {author} {\bibfnamefont
			{R.}~\bibnamefont {Camacho-Morales}}, \bibinfo {author} {\bibfnamefont
			{Y.}~\bibnamefont {Ma}}, \bibinfo {author} {\bibfnamefont {G.}~\bibnamefont
			{Zhang}}, \bibinfo {author} {\bibfnamefont {M.}~\bibnamefont {Woolley}},
		\bibinfo {author} {\bibfnamefont {D.}~\bibnamefont {Neshev}}, \emph
		{et~al.},\ }\bibfield  {title} {\bibinfo {title} {Dynamic nonlinear image
			tuning through magnetic dipole quasi-BIC ultrathin resonators},\ }\href
	{https://doi.org/10.1002/advs.201802119} {\bibfield  {journal} {\bibinfo
			{journal} {Adv. Sci.}\ }\textbf {\bibinfo {volume} {6}},\ \bibinfo {pages}
		{1802119} (\bibinfo {year} {2019})}\BibitemShut {NoStop}%
	\bibitem [{\citenamefont {Koshelev}\ \emph {et~al.}(2019)\citenamefont
		{Koshelev}, \citenamefont {Tang}, \citenamefont {Li}, \citenamefont {Choi},
		\citenamefont {Li},\ and\ \citenamefont {Kivshar}}]{Koshelev2019}%
	\BibitemOpen
	\bibfield  {author} {\bibinfo {author} {\bibfnamefont {K.}~\bibnamefont
			{Koshelev}}, \bibinfo {author} {\bibfnamefont {Y.}~\bibnamefont {Tang}},
		\bibinfo {author} {\bibfnamefont {K.}~\bibnamefont {Li}}, \bibinfo {author}
		{\bibfnamefont {D.-Y.}\ \bibnamefont {Choi}}, \bibinfo {author}
		{\bibfnamefont {G.}~\bibnamefont {Li}},\ and\ \bibinfo {author}
		{\bibfnamefont {Y.}~\bibnamefont {Kivshar}},\ }\bibfield  {title} {\bibinfo
		{title} {Nonlinear metasurfaces governed by bound states in the continuum},\
	}\href {https://doi.org/10.1021/acsphotonics.9b00700} {\bibfield  {journal}
		{\bibinfo  {journal} {ACS Photonics}\ }\textbf {\bibinfo {volume} {6}},\
		\bibinfo {pages} {1639} (\bibinfo {year} {2019})}\BibitemShut {NoStop}%
	\bibitem [{\citenamefont {Zheng}\ \emph {et~al.}(2022)\citenamefont {Zheng},
		\citenamefont {Xu}, \citenamefont {Huang}, \citenamefont {Smirnova},
		\citenamefont {Hong}, \citenamefont {Ying},\ and\ \citenamefont
		{Rahmani}}]{Zheng2022}%
	\BibitemOpen
	\bibfield  {author} {\bibinfo {author} {\bibfnamefont {Z.}~\bibnamefont
			{Zheng}}, \bibinfo {author} {\bibfnamefont {L.}~\bibnamefont {Xu}}, \bibinfo
		{author} {\bibfnamefont {L.}~\bibnamefont {Huang}}, \bibinfo {author}
		{\bibfnamefont {D.}~\bibnamefont {Smirnova}}, \bibinfo {author}
		{\bibfnamefont {P.}~\bibnamefont {Hong}}, \bibinfo {author} {\bibfnamefont
			{C.}~\bibnamefont {Ying}},\ and\ \bibinfo {author} {\bibfnamefont
			{M.}~\bibnamefont {Rahmani}},\ }\bibfield  {title} {\bibinfo {title}
		{Boosting second-harmonic generation in the LiNbO$_{3}$ metasurface using high-$Q$
			guided resonances and bound states in the continuum},\ }\href
	{https://doi.org/10.1103/physrevb.106.125411} {\bibfield  {journal} {\bibinfo
			{journal} {Phys. Rev. B}\ }\textbf {\bibinfo {volume} {106}},\ \bibinfo
		{pages} {125411} (\bibinfo {year} {2022})}\BibitemShut {NoStop}%
	\bibitem [{\citenamefont {Liu}\ \emph {et~al.}(2023)\citenamefont {Liu},
		\citenamefont {Qu}, \citenamefont {Gu}, \citenamefont {Zhang}, \citenamefont
		{Wu}, \citenamefont {Cai}, \citenamefont {Ren},\ and\ \citenamefont
		{Xu}}]{Liu2023a}%
	\BibitemOpen
	\bibfield  {author} {\bibinfo {author} {\bibfnamefont {Q.}~\bibnamefont
			{Liu}}, \bibinfo {author} {\bibfnamefont {L.}~\bibnamefont {Qu}}, \bibinfo
		{author} {\bibfnamefont {Z.}~\bibnamefont {Gu}}, \bibinfo {author}
		{\bibfnamefont {D.}~\bibnamefont {Zhang}}, \bibinfo {author} {\bibfnamefont
			{W.}~\bibnamefont {Wu}}, \bibinfo {author} {\bibfnamefont {W.}~\bibnamefont
			{Cai}}, \bibinfo {author} {\bibfnamefont {M.}~\bibnamefont {Ren}},\ and\
		\bibinfo {author} {\bibfnamefont {J.}~\bibnamefont {Xu}},\ }\bibfield
	{title} {\bibinfo {title} {Boosting second harmonic generation by merging
			bound states in the continuum},\ }\href
	{https://doi.org/10.1103/physrevb.107.245408} {\bibfield  {journal} {\bibinfo
			{journal} {Phys. Rev. B}\ }\textbf {\bibinfo {volume} {107}},\ \bibinfo
		{pages} {245408} (\bibinfo {year} {2023})}\BibitemShut {NoStop}%
	\bibitem [{\citenamefont {Liu}\ and\ \citenamefont {Zhou}(2023)}]{Liu2023}%
	\BibitemOpen
	\bibfield  {author} {\bibinfo {author} {\bibfnamefont {R.}~\bibnamefont
			{Liu}}\ and\ \bibinfo {author} {\bibfnamefont {C.}~\bibnamefont {Zhou}},\
	}\bibfield  {title} {\bibinfo {title} {Second harmonic generation in an
			anisotropic lithium niobate metasurface governed by quasi-BICs},\ }\href
	{https://doi.org/10.1364/OL.504379} {\bibfield  {journal} {\bibinfo
			{journal} {Opt. Lett.}\ }\textbf {\bibinfo {volume} {48}},\ \bibinfo {pages}
		{6565} (\bibinfo {year} {2023})}\BibitemShut {NoStop}%
	\bibitem [{\citenamefont {Feng}\ \emph {et~al.}(2023)\citenamefont {Feng},
		\citenamefont {Liu}, \citenamefont {Chen}, \citenamefont {Wu},\ and\
		\citenamefont {Xiao}}]{Feng2023}%
	\BibitemOpen
	\bibfield  {author} {\bibinfo {author} {\bibfnamefont {S.}~\bibnamefont
			{Feng}}, \bibinfo {author} {\bibfnamefont {T.}~\bibnamefont {Liu}}, \bibinfo
		{author} {\bibfnamefont {W.}~\bibnamefont {Chen}}, \bibinfo {author}
		{\bibfnamefont {F.}~\bibnamefont {Wu}},\ and\ \bibinfo {author}
		{\bibfnamefont {S.}~\bibnamefont {Xiao}},\ }\bibfield  {title} {\bibinfo
		{title} {Enhanced sum-frequency generation from etchless lithium niobate
			empowered by dual quasi-bound states in the continuum},\ }\href
	{https://doi.org/10.1007/s11433-023-2223-5} {\bibfield  {journal} {\bibinfo
			{journal} {Sci. China Phys. Mech. Astron.}\ }\textbf {\bibinfo {volume}
			{66}},\ \bibinfo {pages} {124214} (\bibinfo {year} {2023})}\BibitemShut
	{NoStop}%
	\bibitem [{\citenamefont {Tu}\ \emph {et~al.}(2024)\citenamefont {Tu},
		\citenamefont {Feng}, \citenamefont {Li}, \citenamefont {Xing}, \citenamefont
		{Wu}, \citenamefont {Liu},\ and\ \citenamefont {Xiao}}]{Tu2024}%
	\BibitemOpen
	\bibfield  {author} {\bibinfo {author} {\bibfnamefont {X.}~\bibnamefont
			{Tu}}, \bibinfo {author} {\bibfnamefont {S.}~\bibnamefont {Feng}}, \bibinfo
		{author} {\bibfnamefont {J.}~\bibnamefont {Li}}, \bibinfo {author}
		{\bibfnamefont {Y.}~\bibnamefont {Xing}}, \bibinfo {author} {\bibfnamefont
			{F.}~\bibnamefont {Wu}}, \bibinfo {author} {\bibfnamefont {T.}~\bibnamefont
			{Liu}},\ and\ \bibinfo {author} {\bibfnamefont {S.}~\bibnamefont {Xiao}},\
	}\bibfield  {title} {\bibinfo {title} {Enhanced second-harmonic generation in
			high-$Q$ all-dielectric metasurfaces with backward frequency conversion},\
	}\href {https://doi.org/10.1103/PhysRevA.109.063522} {\bibfield  {journal}
		{\bibinfo  {journal} {Phys. Rev. A}\ }\textbf {\bibinfo {volume} {109}},\
		\bibinfo {pages} {063522} (\bibinfo {year} {2024})}\BibitemShut {NoStop}%
	\bibitem [{\citenamefont {He}\ and\ \citenamefont {Wang}(2024)}]{He2024a}%
	\BibitemOpen
	\bibfield  {author} {\bibinfo {author} {\bibfnamefont {W.}~\bibnamefont
			{He}}\ and\ \bibinfo {author} {\bibfnamefont {Y.}~\bibnamefont {Wang}},\
	}\bibfield  {title} {\bibinfo {title} {Enhancement of second-harmonic
			generation in a lithium niobate metasurface by exploring the bound states in
			the continuum},\ }\href {https://doi.org/10.1364/oe.538446} {\bibfield
		{journal} {\bibinfo  {journal} {Opt. Express}\ }\textbf {\bibinfo {volume}
			{32}},\ \bibinfo {pages} {39415} (\bibinfo {year} {2024})}\BibitemShut
	{NoStop}%
	\bibitem [{\citenamefont {Liu}\ \emph {et~al.}(2025)\citenamefont {Liu},
		\citenamefont {Qin}, \citenamefont {Qiu}, \citenamefont {Tu}, \citenamefont
		{Qiu}, \citenamefont {Wu}, \citenamefont {Yu}, \citenamefont {Liu},\ and\
		\citenamefont {Xiao}}]{Liu2025}%
	\BibitemOpen
	\bibfield  {author} {\bibinfo {author} {\bibfnamefont {T.}~\bibnamefont
			{Liu}}, \bibinfo {author} {\bibfnamefont {M.}~\bibnamefont {Qin}}, \bibinfo
		{author} {\bibfnamefont {J.}~\bibnamefont {Qiu}}, \bibinfo {author}
		{\bibfnamefont {X.}~\bibnamefont {Tu}}, \bibinfo {author} {\bibfnamefont
			{H.}~\bibnamefont {Qiu}}, \bibinfo {author} {\bibfnamefont {F.}~\bibnamefont
			{Wu}}, \bibinfo {author} {\bibfnamefont {T.}~\bibnamefont {Yu}}, \bibinfo
		{author} {\bibfnamefont {Q.}~\bibnamefont {Liu}},\ and\ \bibinfo {author}
		{\bibfnamefont {S.}~\bibnamefont {Xiao}},\ }\bibfield  {title} {\bibinfo
		{title} {Polarization-independent enhancement of third-harmonic generation
			empowered by doubly degenerate quasi-bound states in the continuum},\ }\href
	{https://doi.org/10.1021/acs.nanolett.5c00146} {\bibfield  {journal}
		{\bibinfo  {journal} {Nano Lett.}\ }\textbf {\bibinfo {volume} {25}},\
		\bibinfo {pages} {3646} (\bibinfo {year} {2025})}\BibitemShut {NoStop}%
	\bibitem [{\citenamefont {Sun}\ \emph {et~al.}(2025)\citenamefont {Sun},
		\citenamefont {Wang}, \citenamefont {Wang}, \citenamefont {Cai},
		\citenamefont {Huang}, \citenamefont {Alù},\ and\ \citenamefont
		{Han}}]{Sun2025}%
	\BibitemOpen
	\bibfield  {author} {\bibinfo {author} {\bibfnamefont {K.}~\bibnamefont
			{Sun}}, \bibinfo {author} {\bibfnamefont {K.}~\bibnamefont {Wang}}, \bibinfo
		{author} {\bibfnamefont {W.}~\bibnamefont {Wang}}, \bibinfo {author}
		{\bibfnamefont {Y.}~\bibnamefont {Cai}}, \bibinfo {author} {\bibfnamefont
			{L.}~\bibnamefont {Huang}}, \bibinfo {author} {\bibfnamefont
			{A.}~\bibnamefont {Alù}},\ and\ \bibinfo {author} {\bibfnamefont
			{Z.}~\bibnamefont {Han}},\ }\bibfield  {title} {\bibinfo {title} {High-$Q$
			photonic flat-band resonances for enhancing third-harmonic generation in
			all-dielectric metasurfaces},\ }\href
	{https://doi.org/10.1016/j.newton.2025.100057} {\bibfield  {journal}
		{\bibinfo  {journal} {Newton}\ ,\ \bibinfo {pages} {100057}} (\bibinfo {year}
		{2025})}\BibitemShut {NoStop}%
	\bibitem [{\citenamefont {Koshelev}\ \emph {et~al.}(2018)\citenamefont
		{Koshelev}, \citenamefont {Lepeshov}, \citenamefont {Liu}, \citenamefont
		{Bogdanov},\ and\ \citenamefont {Kivshar}}]{Koshelev2018}%
	\BibitemOpen
	\bibfield  {author} {\bibinfo {author} {\bibfnamefont {K.}~\bibnamefont
			{Koshelev}}, \bibinfo {author} {\bibfnamefont {S.}~\bibnamefont {Lepeshov}},
		\bibinfo {author} {\bibfnamefont {M.}~\bibnamefont {Liu}}, \bibinfo {author}
		{\bibfnamefont {A.}~\bibnamefont {Bogdanov}},\ and\ \bibinfo {author}
		{\bibfnamefont {Y.}~\bibnamefont {Kivshar}},\ }\bibfield  {title} {\bibinfo
		{title} {Asymmetric metasurfaces with high-$Q$ resonances governed by bound
			states in the continuum},\ }\href
	{https://doi.org/10.1103/physrevlett.121.193903} {\bibfield  {journal}
		{\bibinfo  {journal} {Phys. Rev. Lett.}\ }\textbf {\bibinfo {volume} {121}},\
		\bibinfo {pages} {193903} (\bibinfo {year} {2018})}\BibitemShut {NoStop}%
	\bibitem [{\citenamefont {Wang}\ \emph {et~al.}(2023)\citenamefont {Wang},
		\citenamefont {Srivastava}, \citenamefont {Tan}, \citenamefont {Wang},\ and\
		\citenamefont {Singh}}]{Wang2023}%
	\BibitemOpen
	\bibfield  {author} {\bibinfo {author} {\bibfnamefont {W.}~\bibnamefont
			{Wang}}, \bibinfo {author} {\bibfnamefont {Y.~K.}\ \bibnamefont
			{Srivastava}}, \bibinfo {author} {\bibfnamefont {T.~C.}\ \bibnamefont {Tan}},
		\bibinfo {author} {\bibfnamefont {Z.}~\bibnamefont {Wang}},\ and\ \bibinfo
		{author} {\bibfnamefont {R.}~\bibnamefont {Singh}},\ }\bibfield  {title}
	{\bibinfo {title} {Brillouin zone folding driven bound states in the
			continuum},\ }\href {https://doi.org/10.1038/s41467-023-38367-y} {\bibfield
		{journal} {\bibinfo  {journal} {Nat. Commun.}\ }\textbf {\bibinfo {volume}
			{14}},\ \bibinfo {pages} {2811} (\bibinfo {year} {2023})}\BibitemShut
	{NoStop}%
	\bibitem [{\citenamefont {Adi}\ \emph {et~al.}(2024)\citenamefont {Adi},
		\citenamefont {Rosas}, \citenamefont {Beisenova}, \citenamefont {Biswas},
		\citenamefont {Mei}, \citenamefont {Czaplewski},\ and\ \citenamefont
		{Yesilkoy}}]{Adi2024}%
	\BibitemOpen
	\bibfield  {author} {\bibinfo {author} {\bibfnamefont {W.}~\bibnamefont
			{Adi}}, \bibinfo {author} {\bibfnamefont {S.}~\bibnamefont {Rosas}}, \bibinfo
		{author} {\bibfnamefont {A.}~\bibnamefont {Beisenova}}, \bibinfo {author}
		{\bibfnamefont {S.~K.}\ \bibnamefont {Biswas}}, \bibinfo {author}
		{\bibfnamefont {H.}~\bibnamefont {Mei}}, \bibinfo {author} {\bibfnamefont
			{D.~A.}\ \bibnamefont {Czaplewski}},\ and\ \bibinfo {author} {\bibfnamefont
			{F.}~\bibnamefont {Yesilkoy}},\ }\bibfield  {title} {\bibinfo {title}
		{Trapping light in air with membrane metasurfaces for vibrational strong
			coupling},\ }\href {https://doi.org/10.1038/s41467-024-54284-0} {\bibfield
		{journal} {\bibinfo  {journal} {Nat. Commun.}\ }\textbf {\bibinfo {volume}
			{15}},\ \bibinfo {pages} {10049} (\bibinfo {year} {2024})}\BibitemShut
	{NoStop}%
	\bibitem [{\citenamefont {Wu}\ \emph {et~al.}(2024)\citenamefont {Wu},
		\citenamefont {Qi}, \citenamefont {Qin}, \citenamefont {Luo}, \citenamefont
		{Long}, \citenamefont {Wu}, \citenamefont {Sun}, \citenamefont {Jiang},
		\citenamefont {Liu}, \citenamefont {Xiao},\ and\ \citenamefont
		{Chen}}]{Wu2024}%
	\BibitemOpen
	\bibfield  {author} {\bibinfo {author} {\bibfnamefont {F.}~\bibnamefont
			{Wu}}, \bibinfo {author} {\bibfnamefont {X.}~\bibnamefont {Qi}}, \bibinfo
		{author} {\bibfnamefont {M.}~\bibnamefont {Qin}}, \bibinfo {author}
		{\bibfnamefont {M.}~\bibnamefont {Luo}}, \bibinfo {author} {\bibfnamefont
			{Y.}~\bibnamefont {Long}}, \bibinfo {author} {\bibfnamefont {J.}~\bibnamefont
			{Wu}}, \bibinfo {author} {\bibfnamefont {Y.}~\bibnamefont {Sun}}, \bibinfo
		{author} {\bibfnamefont {H.}~\bibnamefont {Jiang}}, \bibinfo {author}
		{\bibfnamefont {T.}~\bibnamefont {Liu}}, \bibinfo {author} {\bibfnamefont
			{S.}~\bibnamefont {Xiao}},\ and\ \bibinfo {author} {\bibfnamefont
			{H.}~\bibnamefont {Chen}},\ }\bibfield  {title} {\bibinfo {title} {Momentum
			mismatch driven bound states in the continuum and ellipsometric phase
			singularities},\ }\href {https://doi.org/10.1103/physrevb.109.085436}
	{\bibfield  {journal} {\bibinfo  {journal} {Phys. Rev. B}\ }\textbf {\bibinfo
			{volume} {109}},\ \bibinfo {pages} {085436} (\bibinfo {year}
		{2024})}\BibitemShut {NoStop}%
	\bibitem [{\citenamefont {Sun}\ \emph {et~al.}(2024)\citenamefont {Sun},
		\citenamefont {Wang},\ and\ \citenamefont {Han}}]{Sun2024}%
	\BibitemOpen
	\bibfield  {author} {\bibinfo {author} {\bibfnamefont {K.}~\bibnamefont
			{Sun}}, \bibinfo {author} {\bibfnamefont {W.}~\bibnamefont {Wang}},\ and\
		\bibinfo {author} {\bibfnamefont {Z.}~\bibnamefont {Han}},\ }\bibfield
	{title} {\bibinfo {title} {High-$Q$ resonances in periodic photonic
			structures},\ }\href {https://doi.org/10.1103/physrevb.109.085426} {\bibfield
		{journal} {\bibinfo  {journal} {Phys. Rev. B}\ }\textbf {\bibinfo {volume}
			{109}},\ \bibinfo {pages} {085426} (\bibinfo {year} {2024})}\BibitemShut
	{NoStop}%
	\bibitem [{\citenamefont {Yue}\ \emph {et~al.}(2025)\citenamefont {Yue},
		\citenamefont {Xie}, \citenamefont {Ding}, \citenamefont {Zhou},
		\citenamefont {Shen},\ and\ \citenamefont {Wang}}]{Yue2025}%
	\BibitemOpen
	\bibfield  {author} {\bibinfo {author} {\bibfnamefont {L.}~\bibnamefont
			{Yue}}, \bibinfo {author} {\bibfnamefont {P.}~\bibnamefont {Xie}}, \bibinfo
		{author} {\bibfnamefont {Q.}~\bibnamefont {Ding}}, \bibinfo {author}
		{\bibfnamefont {X.}~\bibnamefont {Zhou}}, \bibinfo {author} {\bibfnamefont
			{S.}~\bibnamefont {Shen}},\ and\ \bibinfo {author} {\bibfnamefont
			{W.}~\bibnamefont {Wang}},\ }\bibfield  {title} {\bibinfo {title} {Intrinsic
			strong self-hybrid coupling empowered by brillouin zone folding-induced
			high-$Q$ quasibound modes in van der waals metasurfaces},\ }\href
	{https://doi.org/10.1103/PhysRevB.111.045412} {\bibfield  {journal} {\bibinfo
			{journal} {Phys. Rev. B}\ }\textbf {\bibinfo {volume} {111}},\ \bibinfo
		{pages} {045412} (\bibinfo {year} {2025})}\BibitemShut {NoStop}%
	\bibitem [{\citenamefont {Qin}\ \emph {et~al.}(2025)\citenamefont {Qin},
		\citenamefont {Wu}, \citenamefont {Liu}, \citenamefont {Zhang},\ and\
		\citenamefont {Xiao}}]{Qin2025}%
	\BibitemOpen
	\bibfield  {author} {\bibinfo {author} {\bibfnamefont {M.}~\bibnamefont
			{Qin}}, \bibinfo {author} {\bibfnamefont {F.}~\bibnamefont {Wu}}, \bibinfo
		{author} {\bibfnamefont {T.}~\bibnamefont {Liu}}, \bibinfo {author}
		{\bibfnamefont {D.}~\bibnamefont {Zhang}},\ and\ \bibinfo {author}
		{\bibfnamefont {S.}~\bibnamefont {Xiao}},\ }\bibfield  {title} {\bibinfo
		{title} {Enhanced third-harmonic generation and degenerate four-wave mixing
			in an all-dielectric metasurface via brillouin zone folding induced bound
			states in the continuum},\ }\href
	{https://doi.org/10.1103/PhysRevB.111.035414} {\bibfield  {journal} {\bibinfo
			{journal} {Phys. Rev. B}\ }\textbf {\bibinfo {volume} {111}},\ \bibinfo
		{pages} {035414} (\bibinfo {year} {2025})}\BibitemShut {NoStop}%
	\bibitem [{\citenamefont {Overvig}\ \emph {et~al.}(2020)\citenamefont
		{Overvig}, \citenamefont {Malek}, \citenamefont {Carter}, \citenamefont
		{Shrestha},\ and\ \citenamefont {Yu}}]{Overvig2020}%
	\BibitemOpen
	\bibfield  {author} {\bibinfo {author} {\bibfnamefont {A.~C.}\ \bibnamefont
			{Overvig}}, \bibinfo {author} {\bibfnamefont {S.~C.}\ \bibnamefont {Malek}},
		\bibinfo {author} {\bibfnamefont {M.~J.}\ \bibnamefont {Carter}}, \bibinfo
		{author} {\bibfnamefont {S.}~\bibnamefont {Shrestha}},\ and\ \bibinfo
		{author} {\bibfnamefont {N.}~\bibnamefont {Yu}},\ }\bibfield  {title}
	{\bibinfo {title} {Selection rules for quasibound states in the continuum},\
	}\href {https://doi.org/10.1103/PhysRevB.102.035434} {\bibfield  {journal}
		{\bibinfo  {journal} {Phys. Rev. B}\ }\textbf {\bibinfo {volume} {102}},\
		\bibinfo {pages} {035434} (\bibinfo {year} {2020})}\BibitemShut {NoStop}%
	\bibitem [{\citenamefont {Seres}(2001)}]{Seres2001}%
	\BibitemOpen
	\bibfield  {author} {\bibinfo {author} {\bibfnamefont {J.}~\bibnamefont
			{Seres}},\ }\bibfield  {title} {\bibinfo {title} {Dispersion of second-order
			nonlinear optical coefficient},\ }\href
	{https://doi.org/10.1007/s003400100651} {\bibfield  {journal} {\bibinfo
			{journal} {Appl. Phys. B}\ }\textbf {\bibinfo {volume} {73}},\ \bibinfo
		{pages} {705} (\bibinfo {year} {2001})}\BibitemShut {NoStop}%
	\bibitem [{\citenamefont {Wang}\ \emph {et~al.}(2015)\citenamefont {Wang},
		\citenamefont {Bo}, \citenamefont {Wan}, \citenamefont {Li}, \citenamefont
		{Gao}, \citenamefont {Li}, \citenamefont {Zhang},\ and\ \citenamefont
		{Xu}}]{Wang2015}%
	\BibitemOpen
	\bibfield  {author} {\bibinfo {author} {\bibfnamefont {J.}~\bibnamefont
			{Wang}}, \bibinfo {author} {\bibfnamefont {F.}~\bibnamefont {Bo}}, \bibinfo
		{author} {\bibfnamefont {S.}~\bibnamefont {Wan}}, \bibinfo {author}
		{\bibfnamefont {W.}~\bibnamefont {Li}}, \bibinfo {author} {\bibfnamefont
			{F.}~\bibnamefont {Gao}}, \bibinfo {author} {\bibfnamefont {J.}~\bibnamefont
			{Li}}, \bibinfo {author} {\bibfnamefont {G.}~\bibnamefont {Zhang}},\ and\
		\bibinfo {author} {\bibfnamefont {J.}~\bibnamefont {Xu}},\ }\bibfield
	{title} {\bibinfo {title} {High-$Q$ lithium niobate microdisk resonators on a
			chip for efficient electro-optic modulation},\ }\href
	{https://doi.org/10.1364/OE.23.023072} {\bibfield  {journal} {\bibinfo
			{journal} {Opt. Express}\ }\textbf {\bibinfo {volume} {23}},\ \bibinfo
		{pages} {23072} (\bibinfo {year} {2015})}\BibitemShut {NoStop}%
	\bibitem [{\citenamefont {Li}\ \emph {et~al.}(2019)\citenamefont {Li},
		\citenamefont {Liang}, \citenamefont {Luo}, \citenamefont {He},\ and\
		\citenamefont {Lin}}]{Li2019}%
	\BibitemOpen
	\bibfield  {author} {\bibinfo {author} {\bibfnamefont {M.}~\bibnamefont
			{Li}}, \bibinfo {author} {\bibfnamefont {H.}~\bibnamefont {Liang}}, \bibinfo
		{author} {\bibfnamefont {R.}~\bibnamefont {Luo}}, \bibinfo {author}
		{\bibfnamefont {Y.}~\bibnamefont {He}},\ and\ \bibinfo {author}
		{\bibfnamefont {Q.}~\bibnamefont {Lin}},\ }\bibfield  {title} {\bibinfo
		{title} {High-$Q$ 2D lithium niobate photonic crystal slab nanoresonators},\
	}\href {https://doi.org/10.1002/lpor.201800228} {\bibfield  {journal}
		{\bibinfo  {journal} {Laser Photonics Rev.}\ }\textbf {\bibinfo {volume}
			{13}},\ \bibinfo {pages} {1800228} (\bibinfo {year} {2019})}\BibitemShut
	{NoStop}%
	\bibitem [{\citenamefont {Gao}\ \emph {et~al.}(2021)\citenamefont {Gao},
		\citenamefont {Ren}, \citenamefont {Wu}, \citenamefont {Cai},\ and\
		\citenamefont {Xu}}]{Gao2021}%
	\BibitemOpen
	\bibfield  {author} {\bibinfo {author} {\bibfnamefont {B.}~\bibnamefont
			{Gao}}, \bibinfo {author} {\bibfnamefont {M.}~\bibnamefont {Ren}}, \bibinfo
		{author} {\bibfnamefont {W.}~\bibnamefont {Wu}}, \bibinfo {author}
		{\bibfnamefont {W.}~\bibnamefont {Cai}},\ and\ \bibinfo {author}
		{\bibfnamefont {J.}~\bibnamefont {Xu}},\ }\bibfield  {title} {\bibinfo
		{title} {Electro-optic lithium niobate metasurfaces},\ }\href
	{https://doi.org/10.1007/s11433-021-1668-y} {\bibfield  {journal} {\bibinfo
			{journal} {Sci. China Phys. Mech. Astron.}\ }\textbf {\bibinfo {volume}
			{64}},\ \bibinfo {pages} {240362} (\bibinfo {year} {2021})}\BibitemShut
	{NoStop}%
	\bibitem [{\citenamefont {Zhang}\ \emph {et~al.}(2022)\citenamefont {Zhang},
		\citenamefont {Ma}, \citenamefont {Parry}, \citenamefont {Cai}, \citenamefont
		{Camacho-Morales}, \citenamefont {Xu}, \citenamefont {Neshev},\ and\
		\citenamefont {Sukhorukov}}]{Zhang2022}%
	\BibitemOpen
	\bibfield  {author} {\bibinfo {author} {\bibfnamefont {J.}~\bibnamefont
			{Zhang}}, \bibinfo {author} {\bibfnamefont {J.}~\bibnamefont {Ma}}, \bibinfo
		{author} {\bibfnamefont {M.}~\bibnamefont {Parry}}, \bibinfo {author}
		{\bibfnamefont {M.}~\bibnamefont {Cai}}, \bibinfo {author} {\bibfnamefont
			{R.}~\bibnamefont {Camacho-Morales}}, \bibinfo {author} {\bibfnamefont
			{L.}~\bibnamefont {Xu}}, \bibinfo {author} {\bibfnamefont {D.~N.}\
			\bibnamefont {Neshev}},\ and\ \bibinfo {author} {\bibfnamefont {A.~A.}\
			\bibnamefont {Sukhorukov}},\ }\bibfield  {title} {\bibinfo {title} {Spatially
			entangled photon pairs from lithium niobate nonlocal metasurfaces},\ }\href
	{https://doi.org/10.1126/sciadv.abq4240} {\bibfield  {journal} {\bibinfo
			{journal} {Sci. Adv.}\ }\textbf {\bibinfo {volume} {8}},\ \bibinfo {pages}
		{eabq4240} (\bibinfo {year} {2022})}\BibitemShut {NoStop}%
	\bibitem [{\citenamefont {Ma}\ \emph {et~al.}(2023)\citenamefont {Ma},
		\citenamefont {Zhang}, \citenamefont {Jiang}, \citenamefont {Fan},
		\citenamefont {Parry}, \citenamefont {Neshev},\ and\ \citenamefont
		{Sukhorukov}}]{Ma2023}%
	\BibitemOpen
	\bibfield  {author} {\bibinfo {author} {\bibfnamefont {J.}~\bibnamefont
			{Ma}}, \bibinfo {author} {\bibfnamefont {J.}~\bibnamefont {Zhang}}, \bibinfo
		{author} {\bibfnamefont {Y.}~\bibnamefont {Jiang}}, \bibinfo {author}
		{\bibfnamefont {T.}~\bibnamefont {Fan}}, \bibinfo {author} {\bibfnamefont
			{M.}~\bibnamefont {Parry}}, \bibinfo {author} {\bibfnamefont {D.~N.}\
			\bibnamefont {Neshev}},\ and\ \bibinfo {author} {\bibfnamefont {A.~A.}\
			\bibnamefont {Sukhorukov}},\ }\bibfield  {title} {\bibinfo {title}
		{Polarization engineering of entangled photons from a lithium niobate
			nonlinear metasurface},\ }\href
	{https://doi.org/10.1021/acs.nanolett.3c02055} {\bibfield  {journal}
		{\bibinfo  {journal} {Nano Lett.}\ }\textbf {\bibinfo {volume} {23}},\
		\bibinfo {pages} {8091} (\bibinfo {year} {2023})}\BibitemShut {NoStop}%
	\bibitem [{\citenamefont {Liu}\ \emph {et~al.}(2024)\citenamefont {Liu},
		\citenamefont {Qin}, \citenamefont {Feng}, \citenamefont {Tu}, \citenamefont
		{Guo}, \citenamefont {Wu},\ and\ \citenamefont {Xiao}}]{Liu2024}%
	\BibitemOpen
	\bibfield  {author} {\bibinfo {author} {\bibfnamefont {T.}~\bibnamefont
			{Liu}}, \bibinfo {author} {\bibfnamefont {M.}~\bibnamefont {Qin}}, \bibinfo
		{author} {\bibfnamefont {S.}~\bibnamefont {Feng}}, \bibinfo {author}
		{\bibfnamefont {X.}~\bibnamefont {Tu}}, \bibinfo {author} {\bibfnamefont
			{T.}~\bibnamefont {Guo}}, \bibinfo {author} {\bibfnamefont {F.}~\bibnamefont
			{Wu}},\ and\ \bibinfo {author} {\bibfnamefont {S.}~\bibnamefont {Xiao}},\
	}\bibfield  {title} {\bibinfo {title} {Efficient photon-pair generation
			empowered by dual quasibound states in the continuum},\ }\href
	{https://doi.org/10.1103/physrevb.109.155424} {\bibfield  {journal} {\bibinfo
			{journal} {Phys. Rev. B}\ }\textbf {\bibinfo {volume} {109}},\ \bibinfo
		{pages} {155424} (\bibinfo {year} {2024})}\BibitemShut {NoStop}%
	\bibitem [{\citenamefont {Ma}\ \emph {et~al.}(2024)\citenamefont {Ma},
		\citenamefont {Zhang}, \citenamefont {Horder}, \citenamefont {Sukhorukov},
		\citenamefont {Toth}, \citenamefont {Neshev},\ and\ \citenamefont
		{Aharonovich}}]{Ma2024}%
	\BibitemOpen
	\bibfield  {author} {\bibinfo {author} {\bibfnamefont {J.}~\bibnamefont
			{Ma}}, \bibinfo {author} {\bibfnamefont {J.}~\bibnamefont {Zhang}}, \bibinfo
		{author} {\bibfnamefont {J.}~\bibnamefont {Horder}}, \bibinfo {author}
		{\bibfnamefont {A.~A.}\ \bibnamefont {Sukhorukov}}, \bibinfo {author}
		{\bibfnamefont {M.}~\bibnamefont {Toth}}, \bibinfo {author} {\bibfnamefont
			{D.~N.}\ \bibnamefont {Neshev}},\ and\ \bibinfo {author} {\bibfnamefont
			{I.}~\bibnamefont {Aharonovich}},\ }\bibfield  {title} {\bibinfo {title}
		{Engineering quantum light sources with flat optics},\ }\href
	{https://doi.org/10.1002/adma.202313589} {\bibfield  {journal} {\bibinfo
			{journal} {Adv. Mater.}\ }\textbf {\bibinfo {volume} {36}},\ \bibinfo {pages}
		{2313589} (\bibinfo {year} {2024})}\BibitemShut {NoStop}%
\end{thebibliography}

%

\end{document}